\documentclass[12pt]{article}

\pdfoutput=1

\usepackage[top=80pt,bottom=85pt,left=67pt,right=67pt]{geometry}
\usepackage{amssymb,amsmath,xypic}
\usepackage{graphicx,color,subfigure}
\usepackage{float}
\usepackage{sectsty}
\usepackage{cite}
\usepackage[debug,pageanchor=false]{hyperref}
\definecolor{link}{rgb}{.8,.15,.1}
\hypersetup{colorlinks=true,linkcolor=link,citecolor=link,urlcolor=link,linktocpage}

\makeatletter
\@addtoreset{equation}{section}
\makeatother

\newcommand{\beq}{\begin{equation}}
\newcommand{\eeq}{\end{equation}}
\newcommand{\bea}{\begin{eqnarray}}
\newcommand{\eea}{\end{eqnarray}}
\newcommand{\nn}{\nonumber}

\begin{document}

\begin{titlepage}

\begin{center}

\vskip .5in 
\noindent

{\Large \bf{AdS$_3$ solutions in Massive IIA with small $\mathcal{N}=(4,0)$ supersymmetry}}

\bigskip\medskip

Yolanda Lozano$^{a,}$\footnote{ylozano@uniovi.es},  Niall T. Macpherson$^{b,}$\footnote{nmacpher@sissa.it}, Carlos Nunez$^{c,}$\footnote{c.nunez@swansea.ac.uk}, Anayeli Ramirez$^{a,}$\footnote{anayelam@gmail.com} \\

\bigskip\medskip
{\small

 $a$: Department of Physics, University of Oviedo,
Avda. Federico Garcia Lorca s/n, 33007 Oviedo, Spain
\vskip 3mm
 $b$: SISSA International School for Advanced Studies,
Via Bonomea 265, 34136 Trieste \\
		and \\INFN sezione di Trieste
\vskip 3mm
 $c$: Department of Physics, Swansea University, Swansea SA2 8PP, United Kingdom.}

\vskip .9cm 
     	{\bf Abstract }

\vskip .1in
\end{center}

\noindent
We study AdS$_3\times \text{S}^2$ solutions in massive IIA that preserve small ${\cal N}=(4,0)$ supersymmetry in terms of an SU(2)-structure on the remaining internal space. We find two new  classes of solutions that are warped products of the form AdS$_3\times  \text{S}^2\times \text{M}_4\times \mathbb{R}$. For the first, M$_4$=CY$_2$ and we find a generalisation of a D4-D8 system involving possible additional branes. For the second, M$_4$ need only be Kahler, and we find a generalisation of the T-dual of solutions based on D3-branes wrapping curves in the base of an elliptically fibered Calabi-Yau 3-fold. Within these classes we find many new locally compact solutions that are foliations of  AdS$_3\times  \text{S}^2\times\text{CY}_2$ over an interval, bounded by various D brane and O plane behaviours. We comment on how these local solutions may be used as the building blocks of  infinite classes of global solutions glued together with defect branes.  Utilising T-duality we find two new classes of AdS$_3\times \text{S}^3\times \text{M}_4$ solutions in IIB. The first backreacts D5s and KK monopoles on the D1-D5 near horizon. The second is a generalisation of the solutions based on D3-branes wrapping curves in the base of an elliptically fibered CY$_3$ that includes non trivial 3-form flux.

\noindent

\vfill
\eject

\end{titlepage}

\tableofcontents

\section{Introduction}
Two dimensional CFTs play a prominent role in string theory and provide the best arena  to test the AdS/CFT correspondence. The conformal group in two dimensions is infinite dimensional and this makes two dimensional CFTs much more tractable than their higher dimensional counterparts, in some cases even exactly solvable \cite{difrancesco}. In turn, certain AdS$_3$ solutions involve only NS-NS fields, and constitute exactly solvable string theory backgrounds \cite{Maldacena:2000hw,Maldacena:2000kv}.  As such there is clear motivation to study the AdS$_3$/CFT$_2$ duality in as much detail as possible.

The canonical example of AdS$_3$ geometry is the near horizon limit of D1 and D5 branes \cite{Maldacena:1997re} which gives rise to an AdS$_3\times \text{S}^3\times \text{CY}_2$ geometry realising small $\mathcal{N}=(4,4)$ superconformal symmetry. The CFT dual is believed to be the free symmetric product orbifold $\text{Sym}^N(\text{CY}_2)$ for CY$_2=\text{T}^4$ or K3 \cite{Giveon:1998ns}. In recent years there has been renewed interest in its study and strong support for this proposal has been provided\cite{Hampton:2018ygz,Gaberdiel:2018rqv,Eberhardt:2018ouy,Eberhardt:2019qcl,Keller:2019suk}.

Despite this early success, small $\mathcal{N}=(4,0)$ AdS$_3$ solutions in 10 and 11 dimensions are rare in the literature, with known cases mostly following from \cite{Maldacena:1997re} via orbifoldings and/or string dualities. These describe the near horizon limit of D1 and D5 branes intersecting with KK-monopoles \cite{Kutasov:1998zh,Sugawara:1999qp,Larsen:1999dh,Okuyama:2005gq}
or D9-branes  \cite{Douglas:1996uz}. These systems play a prominent role in the  microscopical description of five dimensional black holes \cite{Maldacena:1997de,Vafa:1997gr,Minasian:1999qn,Castro:2008ne,Haghighat:2015ega,Couzens:2019wls}. 2d (4,0) CFTs are also central in the description of self-dual strings in 6d (1,0) CFTs, realised in M and F theory \cite{Haghighat:2013gba,Haghighat:2013tka,Kim:2015gha,Gadde:2015tra,Lawrie:2016axq,Couzens:2017way}. The AdS$_3$ duals of D3-branes wrapped on complex curves in F-theory have been recently constructed in \cite{Couzens:2017way}, and will play an important role in this work. More general (4,0) 2d CFTs such as the ones described by the quivers constructed in  \cite{Haghighat:2013gba,Haghighat:2013tka,Gadde:2015tra,Hanany:2018hlz}, are however still lacking a holographic description. One of the motivations of this work will be to fill this gap.

This dearth of holographic duals is symptomatic of the limited classification effort aimed at AdS$_3$ in general\footnote{See however \cite{Deger:2019tem} for some systematic work addressing this from a 3d gauged supergravity perspective.}, which mostly focuses on different superconformal algebras and restrictive  ans$\ddot{\text{a}}$tze (see for instance \cite{Kim:2005ez,Gauntlett:2006af,Gauntlett:2006ns,Kim:2012ek,Kelekci:2016uqv,Eberhardt:2017uup,Couzens:2017nnr,Dibitetto:2018ftj,Macpherson:2018mif}). Bucking this trend are \cite{Kim:2007hv,Colgain:2010wb} and \cite{Couzens:2017way} which do study small $\mathcal{N}=(4,0)$ solutions in M-theory and IIB respectively,  though they still take a restricted ansatz for the fluxes. In this work we shall focus on small $\mathcal{N}=(4,0)$ in massive IIA, we will make no restriction on the allowed fluxes, though we will also make some assumptions.

Our approach to finding AdS$_3$ solutions with small $\mathcal{N}=(4,0)$ superconformal symmetry is to construct Killing spinors which manifestly transform in the $(\textbf{2},\mathbf{2}\oplus\overline{\textbf{2}})$ of the bosonic sub-algebra $\mathfrak{sl}(2)\oplus \mathfrak{su}(2)$ - the same as the fermionic generators of the algebra \cite{Fradkin:1992bz}. The first factor is realised by Killing spinors on AdS$_3$ which transform in the \textbf{2} of $\mathfrak{sl}(2)$, while the second is an SU(2) R-symmetry SU(2)$_R$ that suggests a local description of the geometry and fluxes in which this SU(2)$_R$ is realised by a 2-sphere, that we shall assume is round. We then realise the $\textbf{2}\oplus \overline{\textbf{2}}$ representation of SU(2)$_R$ by taking certain products of Killing spinors on S$^2$ and spinors on the internal 5-manifold. The fundamental building block in this construction, previously exploited in  \cite{Macpherson:2016xwk,Macpherson:2017mvu,Apruzzi:2018cvq,DeLuca:2018buk,Legramandi:2018itv}, are the SU(2) doublets one can form from the Killing spinors on $ \text{S}^2$. 

A major advantage of this R-symmetry based approach to constructing spinors on the internal space is that it is possible to show that $\mathcal{N}=(4,0)$ supersymmetry is actually implied by an $\mathcal{N}=1$ sub-sector through the action of SU(2)$_R$. As such we are able to exploit an existing  geometric classification of $\mathcal{N}=1$ AdS$_3$ solutions \cite{Dibitetto:2018ftj} to extract necessary conditions on the geometry and fluxes. Of course there should be  rather a lot of solutions of the form AdS$_3\times  \text{S}^2\times\text{M}_5$, so in this work we will focus on those for which $\text{ M}_5$ supports an SU(2)-structure\footnote{A similar restriction was taken in \cite{Passias:2018zlm} for $\mathcal{N}=2$ AdS$_4$ in massive IIA.}. We do this in part to try and ensure that we realise small $\mathcal{N}=(4,0)$ rather than some larger superconformal algebra which contains this. That SU(2)-structure implies the small algebra and no more is certainly not a theorem but experience suggests to us that bigger algebras that contain this (such as large $\mathcal{N}=(4,0)$) will require an Identity-structure on M$_5$. Another reason to focus on SU(2)-structure is to keep this work focused, and leave the more generic case with $\text{ M}_5$ supporting an identity structure for the future.\\
~\\
The layout of the paper is as follows: 

In section \ref{sec:two} (with supplementary material in appendix \ref{sec:app}) we perform the technical ground work of constructing spinors transforming in the $\textbf{2}\oplus \overline{\textbf{2}}$ representation of SU(2)$_R$, and extracting necessary and sufficient conditions on the geometry and fluxes for a solution to realise small $\mathcal{N}=(4,0)$. We find two classes of solutions.

Class I is a warped product of the form AdS$_3\times  \text{S}^2\times \text{CY}_2\times \mathbb{R}$ that we study in section \ref{sec: Class I}. The class is summarised and derived in sections \ref{sec:classIsummery} and \ref{sec:classIderivation} respectively. In section \ref{sec:D5CY2} (up to T-duality) we find a generalisation of the D1-D5 near horizon with source D5 branes back reacted on CY$_2$. In section \ref{eq:classICYfoliations} we find several new compact local solutions in massive IIA that are foliations of AdS$_3\times  \text{S}^2\times \text{CY}_2$ over a bounded interval.

Class II is a warped product of the form AdS$_3\times  \text{S}^2\times \text{M}_4\times \mathbb{R}$, where $\text{M}_4$ is now a Kahler  four-manifold. The class is summarised in section \ref{sec:classIIsummery}  and derived in section \ref{sec:classIIderivation}. Exploiting T-duality, in section \ref{eq:D3curves}, we find a generalisation of the class of D3 branes wrapping curves in the base of an elliptically fibered CY$_3$ \cite{Couzens:2017way} - with non trival 3-form flux turned on. In section \ref{eq:classIICYfoliations} we then find further local  AdS$_3\times  \text{S}^2\times \text{CY}_2$ foliations that are compact.

In section \ref{sec:calibrations} we establish that the local solutions found in sections \ref{eq:classICYfoliations} and \ref{eq:classIICYfoliations} may be used to construct a significantly richer variety of globally compact solutions by using defect branes to glue local solutions together. 

Finally in section \ref{summery} we summarise and discuss some future directions.

\section{AdS$_3\times  \text{S}^2$ solutions with small $\mathcal{N}=(4,0)$ supersymmetry and SU(2)-structure}\label{sec:two}
In this section we derive geometric conditions for a class of warped AdS$_3$ solutions preserving small $\mathcal{N}=(4,0)$ supersymmetry in massive IIA.\\ 
~\\
The small $\mathcal{N}=(4,0)$ super-conformal algebra contains a bosonic sub-algebra
\beq\label{eq:subalgebra}
\mathfrak{sl}(2)\oplus \mathfrak{su}(2)
\eeq    
that should be realised geometrically by the solutions we are interested in. The $\mathfrak{sl}(2)$ factor is simply realised by AdS$_3$. The $\mathfrak{su}(2)$ factor is an R-symmetry, we shall denote SU(2)$_R$, that should be realised by the 7 dimensional internal space M$_7$. This indicates that M$_7$ should admit a local description that contains a 2-sphere, that may be round or appear as part of an SU(2)$\times$U(1) preserving squashed 3-sphere, foliated over the remaining directions. In this work we shall assume the former and seek solutions with metric decomposing as
\beq\label{eq:met}
ds^2= e^{2A} ds^2(\text{AdS}_3)+ ds^2(\text{M}_7),~~~~ds^2(\text{M}_7)=e^{2C} ds^2( \text{S}^2)+ ds^2(\text{M}_5),
\eeq
where the warp factors $e^{2A},e^{2C}$ and dilaton $\Phi$ have support in M$_5$, and the fluxes depend on the AdS$_3$ and $ \text{S}^2$ directions only through their respective volume forms vol(AdS$_3$) and vol(S$^2$). This is sufficient to ensure that we respect the isometries of AdS$_3$ and $ \text{S}^2$. However to guarantee small $\mathcal{N}=(4,0)$ supersymmetry we must solve the supersymmetry constraints. Our strategy to achieve this is as follows
\begin{enumerate}
\item Construct spinors on that transform in the $(\textbf{2},\textbf{2}\oplus \overline{\textbf{2}})$ representation of SL(2)$\times$SU(2), ensuring consistency with the bosonic sub-algebra of small $\mathcal{N}=(4,0)$ superconformal symmetry.
\item Reduce our considerations to an $\mathcal{N}=1$ sub-sector of this spinor that manifestly implies $\mathcal{N}=(4,0)$ through the action of the R-symmetry - this requires the bosonic supergravity fields to be SL(2)$\times$SU(2) singlets.
\item Exploit an existing $\mathcal{N}=1$ AdS$_3$ classification \cite{Dibitetto:2018ftj} to obtain sufficient conditions on the geometry and fluxes for a solution with small $\mathcal{N}=(4,0)$ and SU(2)-structure in IIA to exist.
\item Study the classes consistent with our assumptions, and simplify them as much as possible in a coordinate patch away from the loci of possible sources.  
\end{enumerate}
We will deal with points 1-2 in section \ref{sec:su2r}, which is the most technical part of the paper and can be skipped on a first reading. Section \ref{sec:geometricsusy} deals with point 3. For point 4 there are 2 classes of solutions to study with SU(2)-structure, specialised to conformal Calabi-Yau and Kahler structure. We present these and study them in sections  \ref{sec: Class I} and \ref{sec: Class II}. Those readers merely interested in the results can find summaries of these classes in sections \ref{sec:classIsummery} and \ref{sec:classIIsummery}.\\
~\\
Following 1-4 leads to necessary and sufficient conditions for two classes of solutions to exist in the absence of localised sources. When these are present, the derivation is still completely valid away from their loci, but at these specific points we must solve some additional constraints.  Namely that the source corrected Bianchi identities hold and that the sources have a supersymmetric embedding - ie they must be calibrated \cite{Lust:2010by,Legramandi:2018qkr}. We shall come back to this issue in section \ref{sec:calibrations}. However, from a practical perspective one should appreciate that it is often not necessary to check these conditions explicitly. In particular, if the warp factors and relevant parts of the fluxes reproduce the behaviour of known localised supersymmetric sources (ie branes, O-planes and their generalisations) at some point in the geometry, then one knows that these additional conditions must follow. We will exploit this fact in sections \ref{eq:classICYfoliations} and \ref{eq:classIICYfoliations}.\\
~\\
In the next section we shall construct $\mathcal{N}=(4,0)$ spinors that manifestly transform under the action of SU(2). We shall then be able to identify an $\mathcal{N}=(1,0)$ sub-sector, which when solved, implies the full $\mathcal{N}=(4,0)$ under the action of SU(2)$_R$.

\subsection{Realising an SU(2) R-symmetry}\label{sec:su2r}
Supersymmetric solutions of type II supergravity come equipped with associated Majorana-Weyl Killing spinors $\epsilon_1,\epsilon_2$, that ensure the vanishing of the dilatino and gravitino variations. As we seek solutions with an AdS$_3$ factor that preserve $\mathcal{N}=(4,0)$ supersymmetry we can decompose these spinors as
\beq\label{eq10dspinors}
\epsilon_1=\sum_{I=1}^4 \zeta^I \otimes v_+\otimes \chi^I_1,~~~~\epsilon_2= \sum_{I=1}^4\zeta^I \otimes v_{\mp}\otimes \chi^I_2 
\eeq
where $\zeta^I$ are 4 independent Majorana Killing spinors on unit radius AdS$_3$ and  $\chi^I_{1,2}$  each contain 4 independent Majorana spinors on M$_7$. The remaining factors  $v_{\pm}$ are auxiliary vectors that are required to make $\epsilon_{1,2}\in \text{Cliff}(1,9)$ as we decompose in terms of spinors in 3 and 7 dimensions. They also take care of 10 dimensional chirality, so the upper/lower signs are taken in IIA/B. The 10 dimensional gamma matrices undergo a  similar decomposition as
\beq
\Gamma_{M}=e^{A}\gamma^{(3)}_M\otimes \sigma_3\otimes \mathbb{I},~~~~\Gamma_A= \mathbb{I}\otimes \sigma_2\otimes\gamma^{(7)}_A
\eeq
where $\gamma^{(3)}_M$ are real and defined on unit radius AdS$_3$, and $\gamma^{(7)}_A$ are defined on M$_7$. $ \sigma_i$ are the Pauli matrices so the 10 dimensional chirality matrix is $\hat\gamma=\mathbb{I}\otimes \sigma_1\otimes \mathbb{I}$, so that $\sigma_1 v_{\pm}=\pm v_{\pm}$. The intertwiner, defining Majorana conjugation as $\epsilon^c= B^{(10)}\epsilon^*$, is $B^{(10)}= \mathbb{I}\otimes\mathbb{I}\otimes B^{(7)}$ so that $v_{\pm}$ are real and $B^{(7)-1} \gamma^{(7)}_A B^{(7)}=-\gamma^{(7)*}_A$, $B^{(7)}B^{(7)*}=1$. 

There are actually several distinct types of superconformal algebras corresponding to $\mathcal{N}=(4,0)$ (see \cite{Beck:2017wpm} for a classification). One way to ensure that we have (at least\footnote{The small $\mathcal{N}=(4,0)$ superalgebra is a sub-algebra of several larger ones - notably the large $\mathcal{N}=(4,0)$ superalgebra $\mathfrak{D}(2,1,\alpha)$ \cite{Sevrin:1988ew}. We will not be concerned with this subtlety in this paper.})  small $\mathcal{N}=(4,0)$ is to demand that the internal parts of our $\mathcal{N}=(4,0)$ spinors are charged under an SU(2) R-symmetry, specifically transforming in the $\textbf{2}\oplus \overline{\textbf{2}}$ representation. Then since the spinors on AdS$_3$ are charged under SL(2) we manifestly realise the bosonic sub-algebra of small $\mathcal{N}=(4,0)$ superconformal symmetry \eqref{eq:subalgebra}. If the internal spinors are charged under SU(2)$_R$ it should be possible to construct a $\chi^I_{1,2}$ realising a 4d basis of the SU(2) Lie algebra $\frac{i}{2}\Sigma_i$,  when acted on by the spinoral Lie derivative - ie
\beq\label{eq:4dsu2}
\mathcal{L}_{K_i}\chi_{1,2}^I= \frac{i}{2}(\Sigma_i)^I_{~J} \chi_{1,2}^J
\eeq
where  $K_i$ are the 3 Killing vectors of SU(2). Let us now construct such SU(2) spinors.

As we decompose the internal space as M$_7= \text{S}^2\times \text{M}_5$, we anticipate that the Killing spinors on $ \text{S}^2$ will realise SU(2)$_R$.  For a unit norm 2-sphere, the Killing spinors $\xi$ can be taken to obey
\beq
\nabla^{ \text{S}^2}_{\mu}\xi=\frac{i}{2} \sigma_{\mu}\xi,~~~~|\xi|^2=1,
\eeq
where $\mu$ are flat indices on the unit sphere and $\sigma_{\mu}$ are the first 2 Pauli matrices. The chirality matrix is $\sigma_3$ and Majorana conjugation is defined as $\xi^c= \sigma_2 \xi^*$. To incorporate this into M$_7$ we further decompose the gamma matrices as
\beq\label{eq:7dgammas}
\gamma^{(7)}_{\mu}= e^{C} \sigma_{\mu}\otimes \mathbb{I},\quad\gamma^{(7)}_a = \sigma_3\otimes \gamma_a,\quad B^{(7)}= \sigma_2\otimes B,
\eeq
with $\gamma_a$ gamma matrices in 5d and $BB^*=-1$, $B^{-1}\gamma_a B=\gamma_a^*$.
As established in \cite{Macpherson:2016xwk}, the $ \text{S}^2$ Killing spinors so defined may be used to construct two independent SU(2) doublets
\beq\label{eq:spinordoubletSU2}
\xi^{\alpha} = \left(\begin{array}{c}\xi\\ \xi^c\end{array}\right)^{\alpha},\quad\quad \hat\xi^{\alpha} = \left(\begin{array}{c}i\sigma_3\xi\\ i\sigma_3\xi^c\end{array}\right)^{\alpha},
\eeq
that transform under SU(2) as
\beq\label{eq:su2lie1}
{\cal L}_{K_i}\xi^{\alpha} = \frac{i}{2} (\sigma_i)^{\alpha}_{~\beta} \xi^{\beta},\quad\quad {\cal L}_{K_i}\hat\xi^{\alpha} = \frac{i}{2} (\sigma_i)^{\alpha}_{~\beta} \hat\xi^{\beta},
\eeq
with $\frac{i}{2}\sigma_i$ a 2d representation of the SU(2) Lie algebra and where the 1-forms dual to the Killing vectors can now be taken to be
\beq\label{eq:su21forms}
K_i= \epsilon_{ijk}y_jdy_k,
\eeq
for $y_i$ embedding coordinates on the unit 2-sphere.
One can form 7 dimensional spinors giving rise to a 4 dimensional representation of SU(2) in terms of a spinor on M$_5$,  $\eta$, that is an SU(2) singlet, with which one defines
\beq\label{eq:douletsinglet}
\eta^{\alpha} = \left(\begin{array}{c}\eta\\ \eta^c\end{array}\right)^{\alpha}.
\eeq
One can then contract the $ \text{S}^2$ and M$_5$ doublets to form a 7-dimensional SU(2) spinor
\beq\label{eq:7dsu2spinor}
\chi^I= {\cal M}^I_{{\alpha}\beta}\xi^{\alpha}\otimes \eta^{\beta},~~~~{\cal M}^I=(\sigma_2\sigma_1,\sigma_2\sigma_2,\sigma_2\sigma_3,-i \sigma_2)^I,  
\eeq
where all components of $\chi^I$ are Majorana\footnote{Demanding this actually fixes the second component of \eqref{eq:douletsinglet} in terms of the first. Note that the form of \eqref{eq:7dsu2spinor} is very similar to that of the SO(4) spinors constructed in \cite{Macpherson:2018mif}.}.
Using \eqref{eq:su2lie1} and Pauli matrix identities it is not hard to show that $\chi^I$ transforms as \eqref{eq:4dsu2} with specific 4 dimensional representation
\beq
\Sigma_i = (\sigma_2\otimes \sigma_1,-\sigma_2\otimes \sigma_3,~\mathbb{I}\otimes \sigma_2)_i,
\eeq
which is equivalent to the $\textbf{2}\oplus\overline{\textbf{2}}$ representation of SU(2)\footnote{Specifically the similarity transformation
\beq
S\sim\left(\begin{array}{cccc}0&0& s&-i s\\ i \overline{s}&-\overline{s}&0&0\\0&0& s& is\\i \overline{s}&\overline{s}&0&0\end{array}\right), 
\eeq
for $s= e^{i \frac{\pi}{4}}$ is such that $\frac{i}{2}S\Sigma_i S^{-1}= \frac{i}{2}\sigma_i \oplus (\frac{i}{2}\sigma_i)^{*}$}.
Since $ \text{S}^2$ only preserves 2 supercharges, it is perhaps not obvious that \eqref{eq:7dsu2spinor} will give rise to 4. However, since \eqref{eq10dspinors} couples the 7 dimensional SU(2) spinors to 4 independent AdS$_3$ spinors, this is guaranteed as long as the components of $\chi^I$ are independent - making use of appendix \ref{sec:app} it is not hard to establish that
\beq
\chi^{I\dag}\chi^J = |\eta|^2\delta^{IJ}
\eeq
which confirms this. Let us stress that although we used $\xi^{\alpha}$ to form $\chi^I$, we can also use $\hat\xi^{\alpha}$, which gives a further 7 dimensional
 SU(2) spinor independent of the first.

The most general expressions we can write for the 7 dimensional SU(2) charged  factors of  \eqref{eq10dspinors} are then
\beq\label{eq:genSU2spinor}
\chi_1^I= \frac{1}{\sqrt{2}}e^{\frac{A}{2}}{\cal M}^I_{\alpha\beta}\big(\xi^{\alpha}\otimes \eta^{\beta}_1+\hat\xi^{\alpha}\otimes\hat \eta^{\beta}_1\big),~~~~~~~~
\chi_2^I= \frac{1}{\sqrt{2}}e^{\frac{A}{2}}{\cal M}^I_{\alpha\beta}\big(\xi^{\alpha}\otimes \eta^{\beta}_2+\hat\xi^{\alpha}\otimes \hat\eta^{\beta}_2\big),
\eeq
where we introduced 4 spinors on M$_5$ $(\eta_1,\hat\eta_1,\eta_2,\hat\eta_2)$. The 10 dimensional spinors of \eqref{eq10dspinors} contain 4 independent $\mathcal{N}=1$ sub-sectors, ie each term in the sums. Because the solutions we seek have an AdS$_3$ factor, $d=10$ supersymmetry is implied by 4 sets of reduced $d=7$ conditions - 1 for each component of $(\chi_1^I,\chi_2^I)$. As such, each component of the internal spinors is such that \cite{Dibitetto:2018ftj}
\beq\label{eq:norms}
e^{\mp A}|\chi_1|^2\pm |\chi_{2}|^2=c_{\pm},
\eeq
for $c_{\pm}$ constant. This relates the norms of these components to the AdS$_3$ warp factor, in such a way that the later can only be an SU(2) singlet if the former are. Setting the charged parts of $|\chi_{1,2}^I|$ to zero imposes the following conditions on the 5d spinors  
\beq\label{eq:5dspincond1}
\hat\eta_{1}^{c\dag}\eta_{1}=\text{Im}(\hat\eta_{1}^{\dag}\eta_{1})=\hat\eta_{2}^{c\dag}\eta_{2}=\text{Im}(\hat\eta_{2}^{\dag}\eta_{2})=0,
\eeq
for the $ \text{S}^2$ zero form bi-linears that give rise to these charged terms (see appendix \ref{sec:app}). In what follows we will fix $c_-=0$ as this is requirement for non zero Romans mass. We can then take $c_+=2$ without loss of generality.  As such the 5d spinors should also obey
\beq\label{eq:5dspincond2}
|\eta_1|^2+|\hat\eta_1|^2=|\eta_2|^2+|\hat\eta_2|^2= 1.
\eeq
There is one final property of the SU(2) spinors we have constructed which it is important to stress. The 4 independent $\mathcal{N}=1$ sub-sectors contained in \eqref{eq:genSU2spinor} each couple to the same spinors in 5 dimensions, and the action of SU(2)$_R$ in \eqref{eq:4dsu2} provides a map between each sub-sector. Specifically, one can write $\chi_{1,2}^I$ in terms of a single component and its  spinoral Lie derivative 
\beq
 \chi_{1,2}^I=\left(\begin{array}{c}\chi_{1,2}\\
 2\mathcal{L}_{K_3}\chi_{1,2}\\-2\mathcal{L}_{K_2}\chi_{1,2}\\2\mathcal{L}_{K_1}\chi_{1,2}\end{array}\right)^I.
\eeq
As such, the $\mathcal{N}=1$ Killing spinor equations following from each of $ \chi_{1,2}^2, \chi_{1,2}^3, \chi_{1,2}^4$  are implied by  $\chi_{1,2}^1$ whenever $\mathcal{L}_{K_i}$ commutes with the dilatino and gravitino variations. This is guaranteed by imposing that all bosonic supergravity fields are singlets under SU(2)$_R$\footnote{The proof is analogous to that in Appendix B of \cite{Macpherson:2018mif}.}. Thus it is sufficient to solve for the $\mathcal{N}=1$ sub-sector involving just $\chi^1_{1,2}$ to know that $\mathcal{N}=(4,0)$ is realised by a solution\footnote{As a redundant check we also performed the analysis of section \ref{sec:geometricsusy} for the other 3 $\mathcal{N}=1$ sub-sectors in $\chi^I_{1,2}$. All that changes is some  signs in the components of the charged SU(2) forms on S$^2$ ($y_i,K_i$ etc) as they appear in \eqref{eq:specificforms1}-\eqref{eq:specificforms2} - no signs changes happen for the SU(2) singlet terms. After factoring out the S$^2$ data one is left with the same necessary and sufficient conditions in 5 dimensions irrespective of which $\mathcal{N}=1$ sub-sector you start with - as expected.}.

Clearly, there should  be rather a lot of distinct classes of solutions consistent with AdS$_3\times  \text{S}^2$. In particular, while \eqref{eq:5dspincond1}-\eqref{eq:5dspincond2} do constrain the 5 dimensional spinors somewhat, they will still lead to many branching possibilities, many of which will have superconformal algebras for which small $\mathcal{N}=(4,0)$ is only a subgroup. To mitigate this issue, for the rest of this paper we will constrain our focus to the particular case where M$_5$ supports an SU(2)-structure - rather than an identity-structure as would be the case generically. We also focus on IIA, leaving IIB for future work. 

In the next section we derive necessary and sufficient geometric conditions for  supersymmetry when M$_5$ supports an SU(2)-structure.

\subsection{Geometric conditions for supersymmetry}\label{sec:geometricsusy}
In the previous section we constructed spinors realising $\mathcal{N}=(4,0)$ and an SU(2) R-symmetry. We further argued that it is sufficient to solve for an $\mathcal{N}=1$ sub-sector, as the rest of the $\mathcal{N}=(4,0)$  conditions are implied by this through the action of SU(2)$_R$, provided that the bosonic fields are SU(2) singlets. In this section we will derive necessary and sufficient conditions for supersymmetry in IIA under the assumption that M$_5$ supports an SU(2)-structure. We shall thus take our $\mathcal{N}=1$ sub-sector to be
\begin{align}
\chi_1&= \frac{e^{\frac{A}{2}}}{\sqrt{2}}(\sigma_2\sigma_1)_{\alpha\beta}\bigg(\sin\left(\frac{\alpha_1+\alpha_2}{2}\right)\xi^{\alpha}+\cos\left(\frac{\alpha_1+\alpha_2}{2}\right)\hat\xi^{\alpha}\bigg)\otimes \eta^{\beta}_1\nn\\[2mm]
\chi_2&= \frac{e^{\frac{A}{2}}}{\sqrt{2}}(\sigma_2\sigma_1)_{\alpha\beta}\bigg(\sin\left(\frac{\alpha_1-\alpha_2}{2}\right)\xi^{\alpha}+\cos\left(\frac{\alpha_1-\alpha_2}{2}\right)\hat\xi^{\alpha}\bigg)\otimes \eta^{\beta}_2,\label{eq:neq1spinor}
\end{align}
with $\alpha_{1,2}$ functions on M$_5$, and where
\beq\label{eq:indedidual5dsinors}
\eta_1= \eta,~~~~\eta_2= e^{i\beta}\eta,~~~~ |\eta|^2=1,
\eeq
with $\beta$ another function on M$_5$. This is the most general parametrisation solving \eqref{eq:norms}-\eqref{eq:5dspincond2} that gives rise to an SU(2)-structure\footnote{Actually one could take $\eta_2=a \eta+ b \eta^c$ with $|a|^2+|b|^2=1$ and still achieve this. However when one plugs this ansatz into the supersymmetry conditions it eventually becomes clear that when $(\text{Re}b,\text{Im}b,\text{Im}a)$ are expressed in polar coordinates all the angles must be constant. They can then be set to any value with rotations of $y_i$ and the vielbein on M$_5$. One can use this freedom to fix $b=0$ and $|a|=1$ without loss of generality. We suppress this subtly.}.

Geometric conditions that imply $\mathcal{N}=1$ for warped AdS$_3$ solutions were recently derived in \cite{Dibitetto:2018ftj}. These are given in terms of a bi- spinor (that is mapped to a poly-form under the Clifford map) constructed from a pair of Majorana spinors $(\chi_1,~\chi_2)$  defined on the internal M$_7$ as
\beq\label{eq:bispinordef}
\Psi_++ i\Psi_- = \chi^{1}\otimes \chi^{2\dag}= \frac{1}{8}\sum_{n=0}^7\frac{1}{n!}\chi^{\dag}_2\gamma^{(7)}_{a_n,...,a_1}\chi_1 dx^{a_1}\wedge...\wedge dx^{a_n},
\eeq
where $\Psi_{\pm}$ are real poly-forms of even/odd degree. Under the assumption of equal internal spinor norm, one has $|\chi_1|^2=|\chi_2|^2=e^{A}$ and the NS 3-form has no electric component. In turn, the RR flux can be expressed as a poly-form
\beq
F= f+ e^{2A}\text{vol}(\text{AdS}_3)\wedge\star_7\lambda(f)
\eeq
with $f$ the sum of the magnetic components of the democratic fluxes. Supersymmetry for unit radius AdS$_3$ in type IIA is then implied by the following geometric conditions
\begin{subequations}
\begin{align}
&d_H(e^{A-\Phi}\Psi_{-})=0,\label{eq:susycond7d1}\\[2mm]
&d_H(e^{2A-\Phi}\Psi_{+})-2  e^{A-\Phi}\Psi_{-}= \frac{e^{3A}}{8}\star_7\lambda(f),\label{eq:susycond7d2}\\[2mm]
&e^{A-\Phi}(f,\Psi_{-})-\frac{1}{2} \text{vol}(\text{M}_7)=0\label{eq:susycond7d3},
\end{align}
\end{subequations}
where $\lambda(X_n)= (-1)^{\frac{n}{2}(n-1)} X_n$ and $(X,Y)$ is the $d=7$ Mukai pairing, defined as $(X,Y)=(\lambda(X)\wedge Y)_7$. The twisted exterior derivative is defined as $d_H=d-H\wedge $. Let us now return to the assumption of equal spinor norm made below \eqref{eq:5dspincond1}. Had we taken 7d spinors with non equal norm instead of \eqref{eq:neq1spinor},  the RHS of \eqref{eq:susycond7d1} would have become $c_- f$ \cite{Dibitetto:2018ftj}. This leads to the necessary condition $f_0 c_-=0$, so a Romans mass is only possible when $c_-=0$ - ie when the spinor norms are equal as in \eqref{eq:neq1spinor}.

Plugging \eqref{eq:neq1spinor} into \eqref{eq:bispinordef} and making use of the bi-linear on $ \text{S}^2$ and M$_5$ defined in appendix \ref{sec:app} it is possible to construct  $\Psi_{\pm}$. However, the completely general expressions for these poly-forms are rather unwieldy. Let us sketch how we simplify them to a more tractable form, by solving some necessary conditions. Upon computing the general form of $\Psi_1$, ie the 1-form part of $\Psi_-$, one finds that it  contains the term
\beq
\Psi_1 = -\frac{1}{8}\cos\alpha_1 \sin\beta K_3+....
\eeq
for $K_i$ the 1-forms dual to the SU(2) Killing vectors defined in \eqref{eq:su21forms}. This term is problematic for \eqref{eq:susycond7d2} as there is no way to generate it under $d$ from the forms that span the $ \text{S}^2$ bi-linears  \eqref{eq:SU2forms}, and making it part of the RR flux would make them charged under SU(2) - thus one necessarily has $\cos\alpha_1 \sin\beta=0$. To determine which factor must vanish one can examine the general form of $\Psi_2$ and $\Psi_3$. In particular the latter can only contain $K_i$ when the former does, due again to \eqref{eq:susycond7d2} and the fact the NS 3-form and RR sector should be SU(2) singlets. We find
\begin{align}
\Psi_2&=-\frac{e^{C}}{8}\sin\alpha_1\sin\beta K_3\wedge V+...,\nn\\[2mm]
\Psi_3&=\frac{e^{C}}{8}\cos\alpha_1(K_1\wedge j_1+K_2\wedge j_2+\cos\beta K_3\wedge j_3)+....,
\end{align} 
where $V$ is a real vector and $(j_1,j_2,j_3)$ are real 2-forms that together span the SU(2)-structure in 5 dimensions, as in appendix \ref{sec:5dbispinors}. As such we fix
\beq
\cos\alpha_1=0,~~~~\alpha_2=\alpha,
\eeq
which we achieve by setting $\alpha_1=\frac{\pi}{2}$ without loss of generality. The 7 dimensional bi-spinors are then given by
\begin{subequations}
\begin{align}
8\Psi_+ &= \big(\sin\alpha + \cos\alpha e^{2C}\text{vol}( \text{S}^2)\big)\wedge\big(y_1 j_1+y_2 j_2-y_3 \text{Im}\psi)\nn\\[2mm]
&+\big(\cos\alpha - \sin\alpha e^{2C}\text{vol}( \text{S}^2)\big)\wedge \text{Re}\psi+- e^{C}\big(K_1\wedge j_1+K_2\wedge j_2-K_3\wedge\text{Im}\psi)\wedge V,\label{eq:specificforms1}\\[2mm]
8\Psi_-&= \big(\cos\alpha-\sin\alpha e^{2C}\text{vol}( \text{S}^2)\big)\wedge \big(y_1 j_1+ y_2 j_2- y_3 \text{Im}\psi\big)\wedge V\nn\\[2mm]
&-\big(\sin\alpha+\cos\alpha e^{2C}\text{vol}( \text{S}^2)\big)\wedge \text{Re}\psi\wedge V-e^{C}\big(dy_1\wedge j_1+dy_2\wedge j_2-dy_3\wedge \text{Im}\psi\big),\label{eq:specificforms2}
\end{align}
\end{subequations}
where to ease presentation we introduced the exponentiated 2-form
\beq
\psi= e^{-i\beta} e^{-i j_3}.
\eeq
$\Psi_{\pm}$ can generically be expressed in terms of an SU(3)-structure in 7 dimensions, but in this case doing so is not particularly illuminating, and \eqref{eq:specificforms1}-\eqref{eq:specificforms2} give far more compact expressions.

We now want to insert \eqref{eq:specificforms1}-\eqref{eq:specificforms2} into \eqref{eq:susycond7d1}-\eqref{eq:susycond7d3}  and derive 5 dimensional conditions that imply these. To do so we decompose
\beq
H= H_3 + e^{2C}H_1\wedge \text{vol}( \text{S}^2),
\eeq
and assume that the RR fluxes only depend on the $ \text{S}^2$ directions through vol$( \text{S}^2)$, and that $(e^A,e^C,e^{\Phi})$ are independent of these directions. Making use of the expressions that map  $(y_i,K_i,\text{vol}( \text{S}^2))$ under the exterior derivative and wedge-product in \eqref{eq:SU2forms}, and  after significant massaging one arrives at necessary and sufficient conditions for supersymmetry. Those independent of the RR forms that follow from \eqref{eq:susycond7d1}-\eqref{eq:susycond7d2} are
\begin{subequations} 
\begin{align}
&2  e^{C}+ \sin\alpha e^A=0,\label{eq:gensusyCOND1}\\[2mm]
&d(e^{3A-\Phi}\sin\alpha\sin\beta)-2e^{2A-\Phi} \cos\alpha \sin\beta V=0,\label{eq:gensusyCOND2}\\[2mm]
&e^{2C}H_1+ \frac{e^A}{2}V- \frac{1}{4}d(e^{2A}\sin\alpha\cos\alpha)=0,\label{eq:gensusyCOND3}\\[2mm]
&d(e^{A-\Phi} \sin\alpha\cos\beta)\wedge V=0,\label{eq:gensusyCOND4}\\[2mm]
&d(e^{3A-\Phi}\sin\alpha \Omega)-2 e^{2A-\Phi}\cos\alpha V\wedge \Omega=0,\label{eq:gensusyCOND5}\\[2mm]
&d(e^{3A-\Phi}\sin\alpha\cos\beta J)-2 e^{2A-\Phi}\cos\alpha\cos\beta V\wedge J- e^{3A-\Phi}\sin\alpha\sin\beta H_3=0,\label{eq:gensusyCOND6}\\[2mm]
&(\sin\beta e^{2A} d(e^{-2A}J)+ \cos\beta H_3)\wedge V=0,\label{eq:gensusyCOND7}\\[2mm]
&\Omega \wedge H_3=(\sin\beta dJ+ \cos\beta H_3)\wedge J=0,\label{eq:gensusyCOND8}
\end{align}
\end{subequations}
where we have repackaged $j_i$ as the more standard SU(2)-structure forms $J,\Omega$
\beq\label{eq:su2conds}
J=j_3,~~~\Omega= j_1+i j_2,~~~J\wedge \Omega=0,~~~~ J\wedge J=\frac{1}{2}\Omega\wedge \overline{\Omega}.
\eeq
From \eqref{eq:susycond7d2} we are also given the following definitions for the RR fluxes
\begin{subequations}
\begin{align}
e^{3A}\star_7 f_6&= d(e^{3A-\Phi} \cos\alpha \cos\beta)+2 e^{2A-\Phi}\sin\alpha \cos\beta V,\label{eq:flux1}\\[2mm]
e^{3A}\star_7 f_4&=(d(e^{3A-\Phi}\cos\alpha \sin\beta J)-2 e^{2A-\Phi}\sin\alpha \sin\beta V\wedge J-e^{3A-\Phi}\cos\alpha \cos\beta H_3)\label{eq:flux2}\\[2mm]
&+ \text{vol}( \text{S}^2)\wedge \big(d(e^{3A+2C-\Phi}\sin\alpha \cos\beta)- e^{3A+2C-\Phi}\cos\alpha\cos\beta H_1 + 2 e^{2A+2C-\Phi}\cos\alpha \cos\beta V\big),\nn\\[2mm]
e^{3A}\star_7 f_2&=-d(\frac{e^{3A-\Phi}}{2}\cos\alpha \cos\beta J\wedge J)- e^{2A-\Phi}\sin\alpha\cos\beta V\wedge J\wedge J+ e^{3A-\Phi}\cos\alpha \sin\beta J\wedge H_3,\nn\\
&+\text{vol}( \text{S}^2)\wedge \bigg(d(e^{3A+2C-\Phi}\sin\alpha\sin\beta J)+ e^{3A+2C-\Phi}\cos\alpha\sin\beta H_1\wedge J\label{eq:flux3}\\
&~~~~~~~~- 2 e^{2A+2C-\Phi}\cos\alpha\sin\beta V\wedge J+ e^{3A+2C-\Phi}\sin\alpha\cos\beta H_3\bigg),\nn\\[2mm]
e^{3A}\star_7 f_0&=-\frac{1}{2}\text{vol}( \text{S}^2)\wedge \bigg(d(e^{3A+2C-\Phi}\sin\alpha\cos\beta  J\wedge J)-2e^{3A+2C-\Phi}\sin\alpha \sin\beta J\wedge H_3\label{eq:flux4}\\
&~~~~~~~~ + e^{3A+2C-\Phi}\cos\alpha \cos\beta H_1\wedge J\wedge J-2  e^{2A+2C-\Phi}\cos\alpha \cos\beta V\wedge J\wedge J \bigg)\nn.
\end{align}
\end{subequations}
Finally \eqref{eq:susycond7d3} gives  the pairing constraints
\begin{align}\label{eq:paring}
e^{A-\Phi}&(f,[(\sin\alpha+\cos\alpha e^{2C}\text{vol}( \text{S}^2))\wedge \text{Re}\psi\wedge V])+2 e^{2C}\text{vol}( \text{S}^2)\wedge V\wedge J\wedge J=0,\nn\\[2mm]
&(f,[\cos\alpha-\sin\alpha e^{2C}\text{vol}( \text{S}^2)\big)\wedge \big(y_1 j_1+ y_2 j_2- y_3 \text{Im}\psi\big)\wedge V])=0.
\end{align}
Equations \eqref{eq:gensusyCOND1}-\eqref{eq:paring} are necessary and sufficient for supersymmetry, but to ensure that we actually have a solution one must impose the Bianchi identities of the magnetic parts of the RR and NS fluxes. Away from localised sources these are
\beq\label{eq:Bianchis}
dH_3=0,~~~~d(e^{2C}H_1)=0,~~~~d_{H}f=0.
\eeq
  In the presence of sources the left hand side of these expressions may be modified by $\delta$-function sources - we shall comment on this when it becomes relevant. Supersymmetry and \eqref{eq:Bianchis} have been shown to imply the remaining equations of motion following from the IIA action \cite{Prins:2013wza}.

The conditions \eqref{eq:gensusyCOND1}-\eqref{eq:gensusyCOND8} contain two physically distinct classes of solutions, namely for $\sin\beta=0$ and $\sin\beta\neq 0$, that we explore in sections \ref{sec: Class I} and \ref{sec: Class II}. To briefly illustrate the difference one can consider \eqref{eq:gensusyCOND2}$\wedge J$ and \eqref{eq:gensusyCOND6}. These may be combined to show in general that
\beq
\sin^2\beta(H_3 - d(\frac{\cos\beta}{\sin \beta}J))=0.
\eeq
When $\sin\beta\neq 0$ this condition gives a unique definition of $H_3$, while when $\sin\beta= 0$ the condition is trivialised and \eqref{eq:gensusyCOND7}-\eqref{eq:gensusyCOND8} merely constrain $H_3$ such that it should give zero when wedged with each of $(J,\Omega,V)$. 

Despite there being two classes, they do contain some common features. We first note that \eqref{eq:gensusyCOND3} defines $e^{2C_1}H_1$ in such a way that its Bianchi identity can only be obeyed away from sources if
\beq
d(e^{A}V)=0.
\eeq
We solve this condition by introducing a local coordinate $\rho$ such that
\beq\label{eq:Vdef}
e^{A}V= d\rho,
\eeq
which enables us to locally decompose the internal 5-manifold as $ds^2(\text{M}_5)= ds^2(\text{M}_4)+e^{-2A}d\rho^2$. The second commonality is \eqref{eq:gensusyCOND1}, which fixes the warp factor of $ \text{S}^2$ uniquely as
\beq
 e^{C}= -\frac{e^A}{2}\sin\alpha.
\eeq
Together these conditions allow us to locally refine the metric ansatz of \eqref{eq:met} as
\beq\label{eq:metrefined}
ds^2= e^{2A}\bigg[ ds^2(\text{AdS}_3)+ \frac{1}{4} \sin^2\alpha ds^2( \text{S}^2)\bigg]+ ds^2(\text{M}_4)+ e^{-2A} d\rho^2,
\eeq
where $\text{M}_4$ supports an SU(2)-structure\footnote{Strictly speaking \eqref{eq:metrefined}  holds in a region of space away from NS sources that do not wrap $ \text{S}^2$. Including such objects is in principle still possible, but they must lie at the intersection of two coordinate patches with local metrics of the form \eqref{eq:metrefined}.}.\\
~\\
Let us now summarise the main results of this section: In section \ref{sec:su2r} we derived general $\mathcal{N}=(4,0)$ spinors on AdS$_3\times  \text{S}^2 \times$ M$_5$ that are manifestly charged under an SU(2) R-symmetry, and compatible with type II supergravity. Any solution consistent with this spinor realises small $\mathcal{N}=(4,0)$ superconformal symmetry. We then established that when one imposes that the physical fields of a solution (metric, dilaton and fluxes) respect SU(2)$_R$, it is sufficient to solve for an $\mathcal{N}=1$ sub-sector of this spinor to know that  $\mathcal{N}=(4,0)$ is realised, as the remaining supercharges are implied by the action of the R-symmetry.  In section  \ref{sec:geometricsusy} we zoomed in on solutions for which M$_5$  supports an SU(2)-structure in massive IIA. We exploited an existing $\mathcal{N}=1$ AdS$_3$ classification of \cite{Dibitetto:2018ftj} to derive necessary and sufficient  conditions on the geometry and fluxes of  an AdS$_3$ solution preserving small $\mathcal{N}=(4,0)$. Finally we established that there are two classes of such solutions, those for which $\sin\beta=0$ and $\sin\beta\neq 0$.\\
~\\
In the next section we study the first class of solutions, where M$_4$ is a conformal Calabi-Yau manifold.

\section{Class I: Conformal Calabi-Yau 2-fold case }\label{sec: Class I}
In this section we study the first class of solutions that follows from the necessary conditions in section \ref{sec:geometricsusy} with $\sin\beta=0$. We find that they are warped products of AdS$_3\times  \text{S}^2\times $CY$_2\times \mathbb{R}$ with all possible massive IIA fluxes turned on.

In section \ref{sec:classIsummery} we present a summary of class I and interpret the types of solutions  that lie within it. In section \ref{sec:classIderivation}, we spell out precisely how class I is derived from the necessary conditions of section \ref{sec:geometricsusy}. Then, in section \ref{sec:D5CY2} we exploit  T-duality to obtain a class of solutions in IIB with D5 branes back reacted on AdS$_3\times \text{S}^3\times$CY$_2$, with $\text{S}^3$ foliated over CY$_2$, that generalises the D1-D5 near horizon. We also show how to realise the sub-class with no fibration as a near horizon limit. Finally, in section \ref{eq:classICYfoliations} we focus on explicit local solutions in massive IIA that are foliations of AdS$_3\times  \text{S}^2\times$CY$_2$ over an interval.

\subsection{Summary of class I}\label{sec:classIsummery}
The solutions of class I have the following NS sector
\begin{align}\label{eq:classI NS}
ds^2&= \frac{u}{\sqrt{h_4 h_8}}\bigg(ds^2(\text{AdS}_3)+\frac{h_8h_4 }{4 h_8h_4+(u')^2}ds^2( \text{S}^2)\bigg)+ \sqrt{\frac{h_4}{h_8}}ds^2(\text{CY}_2)+ \frac{\sqrt{h_4 h_8}}{u} d\rho^2,\\
e^{-\Phi}&= \frac{h_8^{\frac{3}{4}} }{2h_4^{\frac{1}{4}}\sqrt{u}}\sqrt{4h_8 h_4+(u')^2},~~~~ H= \frac{1}{2}d\Bigl(-\rho+\frac{ u u'}{4 h_4 h_8+ (u')^2}\Bigr)\wedge\text{vol}( \text{S}^2)+ \frac{1}{h_8^2}d\rho\wedge H_2.\nn
\end{align}
Here $\Phi$ is the dilaton, $H$ the NS 3-form and $ds^2$ is in string frame. The warping $h_4$ has support on $(\rho,\text{CY}_2)$ while $u$ and $h_8$ have support on $\rho$, with $u'= \partial_{\rho}u$. As shall become clear below, the reason for the notation $h_4,h_8$ is that these functions may be identified with the warp factors of D4 and D8 branes  when $u=1$, the interpretation for generic $u$ is more subtle.\\ 
\\
The  10 dimensional RR fluxes are 
\begin{subequations}
\begin{align}
F_0&=h_8',\label{eq:classIflux1}\\[2mm]
F_2&=-\frac{1}{h_8}H_2-\frac{1}{2}\bigg(h_8- \frac{ h'_8 u'u}{4 h_8 h_4+ (u')^2} \bigg)\text{vol}( \text{S}^2),\label{eq:classIflux2}\\[2mm]
F_4&= \bigg(d\left(\frac{u u'}{2 h_4}\right)+2 h_8  d\rho\bigg) \wedge\text{vol}(\text{AdS}_3)\nn\\[2mm]
& -\frac{h_8}{u} (\hat \star_4 d_4 h_4)\wedge d\rho- \partial_{\rho}h_4\text{vol}(\text{CY}_2)-\frac{u u'}{2 h_8( 4h_8 h_4+ (u')^2)} H_2\wedge \text{vol}( \text{S}^2),\label{eq:classIflux3}
\end{align}
\end{subequations}
with the higher fluxes related to these as $F_6=-\star_{10} F_4,~F_8=\star_{10} F_2,~F_{10}=-\star_{10} F_0$.\\
\\ 
Supersymmetry holds whenever
\beq\label{eq:caseIcon1}
u''=0,~~~~ H_2+ \hat{\star}_4 H_2=0,
\eeq
which makes $u$ a linear function (ie an order 1 polynomial), and where $\hat{\star}_4$ is the Hodge dual on CY$_2$. In a canonical frame on CY$_2$ the associated closed-forms $\hat J,\hat\Omega$ read,
\beq
\hat J=\hat e^1\wedge \hat e^2+\hat e^3\wedge \hat e^4,~~~~\hat \Omega=(\hat e^1+i \hat e^2)\wedge(\hat e^3+ i\hat e^4),
\eeq
and then $H_2$ may be express in terms of 3 arbitrary functions $g_{1,2,3}$ on CY$_2$ as
\beq
H_2= g_1(\hat e^1\wedge \hat e^2- \hat e^3\wedge \hat e^4)+g_2( \hat e^1\wedge \hat e^3+\hat e^2\wedge \hat e^4)+g_3(\hat e^1\wedge \hat e^4-\hat e^2\wedge \hat e^3).
\eeq
The Bianchi identities of the fluxes then impose
\begin{align}\label{eq:BI1}
&h''_8= 0,~~~~dH_2=0\\[2mm]
&\frac{h_8}{u}\nabla^2_{\text{CY}_2}h_4+ \partial_{\rho}^2 h_4 +\frac{2}{h_8^3}(g_1^2+g_2^2+g_3^2)=0,
\nn
\end{align}
away from localised sources.\\
\\
To better understand this class of solutions it is instructive to consider the case with $u=1$ and $g_1=g_2=g_3=0$, so that $H_2=0$. The metric and PDEs of \eqref{eq:BI1} then reduce to those of a D4 brane wrapped on AdS$_3\times  \text{S}^2$ and backreacted on CY$_2$, that is inside the world volume of a D8 wrapped on AdS$_3\times  \text{S}^2\times$CY$_2$. One can compare this to the localised flat space D4-D8 system of \cite{Youm:1999zs} and see that indeed, the warp factors and PDEs match when CY$_2= \mathbb{R}^4$. Of course here there are additional fluxes turned on, but this should be no surprise as what was $\mathbb{R}_{1,4}$ in \cite{Youm:1999zs} has become AdS$_3\times  \text{S}^2$. Thus one should expect additional fluxes to accommodate the fact that this space is no longer flat. More generally, turning on $H_2$ and $u \neq 1$ is essentially a deformation of this system. 

In section \ref{sec:D5CY2} we establish that when one imposes that $\partial_{\rho}$ is an isometry,  class I reduces to the T-dual of D5 branes back reacted on AdS$_3\times \text{S}^3\times \text{CY}_2$ - with $\text{S}^3$ foliated over CY$_2$. It is worth stressing that  class I actually also contains the non-Abelian T-dual of this system as well. To extract this, one can fix 
\beq
u=L \lambda \rho,~~~~h_8= c \rho,~~~~h_4= \frac{\lambda^4}{c}\rho h_5 
\eeq
where $h_5$ depends on CY$_2$ only and should be interpreted as a D5 brane warp factor before the duality. For $h_5=1,H_2=0$ this reproduces the non-Abelian T-dual of the D1-D5 near horizon solution \cite{Sfetsos:2010uq}, which is of course non compact. Class I then provides a general class in which this non compact solution may be embedded. This allows to find a compact completion of this solution in the vein of \cite{Lozano:2016kum,Lozano:2016wrs,Lozano:2017ole,Lozano:2018pcp}, as shown in \cite{LMNR3}.\\
~\\
In the next section we will show how class I is obtained from the necessary supersymmetry conditions derived in section \ref{sec:su2r}.

\subsection{Derivation of class I}\label{sec:classIderivation}
To derive class I we begin by  fixing $\sin\beta=0$. We can in fact fix $\beta=0$ without loss of generality. We begin by refining \eqref{eq:gensusyCOND1}-\eqref{eq:gensusyCOND8} by expanding the exterior derivative in terms of the local coordinate $\rho$ introduced in \eqref{eq:Vdef}, as
\beq
d= d_4 + d\rho\wedge \partial_{\rho}.
\eeq
This reduces \eqref{eq:gensusyCOND5}-\eqref{eq:gensusyCOND6} to
\begin{subequations}
\begin{align}
&d_4(e^{3A-\Phi}\sin\alpha J)=d_4(e^{3A-\Phi}\sin\alpha \Omega)=0,\label{eq:caseIforms1}\\[2mm]
&\partial_{\rho}(e^{3A-\Phi}\sin\alpha J)-2 e^{A-\Phi}\cos\alpha J=\partial_{\rho}(e^{3A-\Phi}\sin\alpha \Omega)-2 e^{A-\Phi}\cos\alpha \Omega=0 \label{eq:caseIforms2}.
\end{align}
\end{subequations}
Using both equations we establish that $d_4(e^{-2A}\cot\alpha)\wedge J=d_4(e^{-2A}\cot\alpha)\wedge \Omega=0$, from which it follows that $d_4(e^{-2A}\cot\alpha)=0$.  We can solve this and \eqref{eq:gensusyCOND4} as
\beq
e^{A-\Phi} \sin\alpha=h_8(\rho),~~~~ e^{-2A}\cot\alpha= \frac{1}{2}\partial_{\rho}\log u(\rho),
\eeq
for $h_8,u$ arbitrary functions. We can then define
\beq
J= e^{-3A+\Phi}\frac{u}{\sin\alpha_2}\hat J,~~~~ \Omega= e^{-3A+\Phi}\frac{u}{\sin\alpha_2}\hat\Omega,
\eeq
which are such that
\beq
d\hat J=d\hat \Omega=0,
\eeq
so that M$_4$ is conformally Calabi-Yau. In turn, the conditions \eqref{eq:gensusyCOND6}-\eqref{eq:gensusyCOND8} constrain $H_3$ as
\beq
H_3=\frac{e^{A}}{h_8^2}V\wedge H_2,~~~~ J\wedge H_2=\Omega\wedge H_2=0,
\eeq
with $H_2$ otherwise free and the factor of $\frac{e^{A}}{h_8^2}$ is chosen for later convenience. A consequence of these conditions is the useful identity $\star_5H_3=-\frac{e^{A}}{h_8^2}H_2$ which holds because the $J\wedge H_2=\Omega\wedge H_2=0$ implies that $H_2$ is anti self dual, and vice versa.

We now turn our attention to the RR fluxes. Using what has been derived thus far, and the fact that
\beq\label{eq:hodgedualofforms}
\star_5 1= \frac{1}{2}V\wedge J\wedge J,~~~~~\star_5 V= \frac{1}{2} J\wedge J,~~~~~\star_5 J= V\wedge J,
\eeq
it is possible to take the Hodge dual of \eqref{eq:flux1}-\eqref{eq:flux4} and arrive at
\begin{subequations}
\begin{align}
f_0 &= h_8'+ \frac{e^{4A} u' u''}{4 u^2},\\[2mm]
f_2&=-\frac{1}{h_8}H_2-\frac{1}{2}\bigg(h_8- \frac{e^{4A} u (u')^2 \left(\frac{h_8}{ u'}\right)'}{4u^2+ e^{4A} (u')^2} \bigg)\text{vol}( \text{S}^2),\\[2mm]
f_4&=-\frac{e^{3A}h_8^2}{u^2} \star_5 d\left(\frac{u^2}{e^{4A} h_8}\right)-\frac{1}{2 h_8}e^{4A}\frac{u u'}{4 u^2+ e^{4A} (u')^2 } H_2\wedge \text{vol}( \text{S}^2)-\frac{e^{4A} h_8 u' u''}{8 u^2}J\wedge J,\\[2mm]
f_6&=\frac{1}{2}\bigg[\frac{e^{7A}h_8^2 u'}{ u(4 u^2+ e^{4A} (u')^2)}\star_5 d\left(\frac{u^2}{e^{4A} h_8}\right)+\frac{1}{2}\left(h_4+ \frac{e^{4A}h_8 u u''}{ u(4 u^2+ e^{4A} (u')^2)}\right)J\wedge J\bigg]\wedge \text{vol}( \text{S}^2).
\end{align}
\end{subequations}
Using these definitions we can now solve \eqref{eq:paring}, which imposes simply
\beq
u''=0.
\eeq
At this point the supersymmetry conditions are completely solved. What remains is to solve the Bianchi identities of the fluxes. Away from localised sources these impose\footnote{In deriving the last of these we make use of the identity
\beq
\frac{h_8^{\frac{5}{4}}}{h_4^{\frac{3}{4}}\sqrt{u}} \star_5 d h_4= \frac{h_8}{u} (\hat \star_4 d_4 h_4)\wedge d\rho+ \partial_{\rho}h_4\text{vol}(\text{CY}_2),
\eeq
and that $\nabla^2_{\text{CY}_2}h_4= \hat\star_4 d_4 \hat\star_4 d_4 h_4$. }
\begin{align}
&h''_8= 0,~~~~dH_2=0\\[2mm]
&\frac{h_8}{u}\nabla^2_{\text{CY}_2}h_4+ \partial_{\rho}^2 h_4 -\frac{1}{h_8^3}\hat\star_4(H_2\wedge H_2)=0.
\nn
\end{align}
Therefore, any solution to \eqref{eq:caseIcon1} and \eqref{eq:BI1} gives a solution in IIA away from localised sources.  When these are included, \eqref{eq:BI1} will have additional $\delta$-function source terms on the LHS and these sources should also be calibrated. We shall return to this in section \ref{sec:calibrations}.

In the next section we will derive a class of solutions with D5-branes backreacted on  AdS$_3\times \text{S}^3\times$CY$_2$, with $\text{S}^3$ fibered over CY$_2$.

\subsection{D5 branes wrapped on $\text{AdS}_3\times \text{S}^3$ and backreacted on CY$_2$}\label{sec:D5CY2}
In this section we derive a class of solutions in IIB with D5 branes and formal KK monopoles that generalises the D1-D5 near horizon.  We begin with the class of solutions in section \ref{sec:classIsummery} and impose that $\partial_{\rho}$ is an isometry. We can achieve this without loss of generality by fixing
\beq
u=(L \lambda)^2c,~~~~h_8= c,~~~~h_4= \lambda^4 c h_5 ,
\eeq
where $h_5$ depends only on the coordinates on CY$_2$ and $(L,\lambda,c)$ are arbitrary constants chosen in this specific combination for convenience. The class then reduces to
\begin{align}
ds^2&= \frac{L^2}{\sqrt{h_5}}\bigg(ds^2(\text{AdS}_3)+\frac{1}{4}ds^2( \text{S}^2)\bigg)+ \lambda^2 \sqrt{h_5}ds^2(\text{CY}_2)+ \frac{\sqrt{h_5}}{L^2} d\rho^2,~~~e^{\Phi}= c^{-\frac{1}{2}}Lh_4^{-\frac{1}{4}},\\[2mm]
 B&= \left(\frac{1}{2}\eta+ \frac{1}{c^2} {\cal A}\right)\wedge d\rho,~~~~F_2=-\frac{1}{c}H_2-\frac{c}{2}\text{vol}( \text{S}^2),~~~F_4= 2 c\text{vol}(\text{AdS}_3)\wedge d\rho-\frac{c\lambda^2}{L^2}\hat\star_4dh_5\wedge d\rho,\nn
\end{align}
where we have introduced 1-form potentials $\eta$ and ${\cal A}$ such that
\beq
d\eta= -\text{vol}( \text{S}^2),~~~~ d{\cal A}= H_2,
\eeq
to write the 2-form NS potential $B$ in a form respecting the isometry $\partial_{\rho}$.\\
~\\
We now T-dualise on $\rho$ (see for instance \cite{Kelekci:2014ima}), which we take to have period $2\pi$, and arrive at the dual IIB solution
\begin{align}\label{eq:classID5s}
ds^2&= \frac{L^2}{\sqrt{h_5}}\bigg[ds^2(\text{AdS}_3)+\frac{1}{4}\bigg(D\psi^2+ds^2( \text{S}^2)\bigg)\bigg]+ \lambda^2 \sqrt{h_5}ds^2(\text{CY}_2),~~~e^{\Phi}= \frac{L^2}{c}h_5^{-\frac{1}{2}},\\[2mm]
F_3&=2 c\bigg[\text{vol}(\text{AdS}_3)+\frac{1}{8}D\psi\wedge\text{vol}( \text{S}^2)\bigg]+\frac{1}{2c }D\psi\wedge d {\cal A}-\frac{c\lambda^2}{L^2}\hat\star_4dh_5,\nn
\end{align}
where $\hat\star_4$ is the Hodge dual on CY$_2$, the NS 3-form and the remaining RR forms are now all trivial and we define 
\beq
D\psi=d\psi+ \eta + \frac{2}{c^2}{\cal A},
\eeq  
for $\partial_{\psi}$ now the isometry direction. Supersymmetry requires that
\beq
d{\cal A}+ \hat\star_4d{\cal A}=0,
\eeq 
where $\hat{J},\hat{\Omega}$ are the 2 and 3 forms on CY$_2$, and the Bianchi identity of the RR 3-form imposes 
\beq\label{eq:refto4point3}
\nabla^2_{\text{CY}_2}h_5-\frac{L^2}{c^4 \lambda^2}\hat\star_4(d{\cal A}\wedge d{\cal A})=0,
\eeq
away from localised sources. When ${\cal A}=0$ and $\psi$ has period $4\pi$ this gives a class of solutions with D5 branes wrapped on AdS$_3\times \text{S}^3$ and backreacted in an arbitrary CY$_2$. Note that one is also free to replace this $\text{S}^3$ by a Lens space, by changing the period of $\psi$, as $\partial_{\psi}$ is an uncharged isometry generically. The effect of turning on ${\cal A}$ is then to formally  place a  Kaluza-Klein monopole into this set-up. The assumption of equal spinor norms made in section \ref{sec:su2r}, means that this class is in fact not the most general one of this type. However solutions in this general class may be reached via SL$(2,\mathbb{R})$ duality\footnote{Equal spinor norms is in 1-to-1 correspondence with having no electric component of the NS 3 form. Such a term, when present, can always be turned off (ie mapped to the RR 3-form) with an SL$(2,\mathbb{R})$ transformation, at which point one necessarily has equal spinor norms.} of \eqref{eq:classID5s}, which will generically turn on $F_1$ and $H_3$ fluxes. In this regard \eqref{eq:classID5s} is a specific duality frame of this more general class.

Clearly  \eqref{eq:classID5s} is closely related to the near horizon limit of coincident D5 and D1 branes that respectively wrap or are smeared on CY$_2$ \cite{Maldacena:1997re,Giveon:1998ns}. In particular if ${\cal A}=0$, $h_5=$ constant and $\psi \sim \psi + 4\pi$, so that there is a round S$^3$, we recover this class and supersymmetry is enhanced to $\mathcal{N}=(4,4)$. Replacing S$^3$ with a Lens space yields the  D1-D5-KK near horizon, which has also been systematically studied \cite{Kutasov:1998zh,Sugawara:1999qp,Larsen:1999dh,Okuyama:2005gq} and preserves only $\mathcal{N}=(4,0)$. Of course when they are non trivial, both ${\cal A}$ and the $\hat\star_4dh_5$ term in $F_3$  break supersymmetry to $\mathcal{N}=(4,0)$ irrespective of the period of $\psi$.  

A viable CFT dual demands a compact internal space which restricts  CY$_2$ to be either $\text{T}^4$ or K3 for the near horizon of D1-D5s and D1-D5-KK. This is no longer the case for generic $h_5$, as the warp factor can cause a non compact space to be restricted to a finite subregion when embedded in 10d. A simple example in this class exhibiting such behaviour was already given in \cite{Macpherson:2018mif},  where a compact solution with D5s and an O5 plane backreacted on AdS$_3\times \text{S}^3 \times \mathbb{R}^4$ was found.

Interestingly, it turns out that the ${\cal A}=0$ limit of \eqref{eq:classID5s} can  also be realised as a near horizon limit of intersecting branes. One begins by compactifying $\mathbb{R}^{1,9}\to \mathbb{R}^{1,5}\times \text{CY}_2$, which preserves $\frac{1}{2}$ of maximal supersymmetry. For the standard D1-D5 system  giving rise to AdS$_3$ in the near horizon, D5 branes would then be wrapped on $\mathbb{R}^{1,1}\times$CY$_2$ and D1 branes placed on $\mathbb{R}^{1,1}$ (smeared across CY$_2$). This breaks supersymmetry to $\frac{1}{4}$ of maximal - enhanced to $\frac{1}{2}$ at the horizon. However, one can also place additional D5s on $\mathbb{R}^{1,1}$ and the common co-dimensions of the other branes at the cost of breaking supersymmetry to $\frac{1}{8}$ of maximal. This leads us to a metric of the form
\beq
ds^2= \frac{1}{\sqrt{H_1 H_5 h_5}} ds^2(\mathbb{R}^{1,1})+\sqrt{\frac{H_5 H_1 }{  h_5}} \big(dr^2+ r^2 ds^2( \text{S}^3)\big)+ \sqrt{\frac{h_5 H_1}{ H_5 }}ds^2(\text{CY}_2)
\eeq
where the warp factors are, respectively,
\beq
H_1=1+ \frac{Q_1}{r^2},~~~~ H_5= 1+ \frac{Q_5}{r^2},~~~~ h_5:~~~ \nabla_{\text{CY}_2}^2 h_5=0,
\eeq
away from the $h_5$ sources. One can then take the near horizon limit of D1s and the D5s corresponding to  $H_5$ by expanding about $r=0$. The metric 
becomes
\beq
ds^2= \frac{1}{\sqrt{h_5}}\bigg[\frac{r^2}{\sqrt{Q_1 Q_5}}ds^2(\mathbb{R}^{1,1})+\sqrt{Q_1 Q_5} \frac{dr^2}{r^2}+ \sqrt{Q_1 Q_5} ds^2( \text{S}^3)\bigg]+ \sqrt{\frac{Q_1}{ Q_5 }}\sqrt{h_5} ds^2(\text{CY}_2),
\eeq
at leading order, which is a solution by itself with supersymmetry  enhanced to $\frac{1}{4}$ of maximal.
Clearly there is an AdS$_3$ factor of radius $(Q_1 Q_5)^{\frac{1}{4}}$, and in fact the entire metric can be easily mapped to that of \eqref{eq:classID5s} (with unit radius AdS$_3$) for ${\cal A}=0$ by  redefining $Q_1,Q_5$ and rescaling $\mathbb{R}_{1,1},h_5$. The same is true of the fluxes, the details of which we have suppressed. Despite how seemingly obvious this near horizon is, as far as the authors are aware, it is absent from the literature. 
Given that a near horizon limit is known, and that the class is relatively simple it would be fruitful to study it in the future.\\
~\\
In the next section we study a class of local solutions in massive IIA following from class I that are a foliation of AdS$_3\times  \text{S}^2\times \text{CY}_2$ over an interval. 
\subsection{Local solutions with AdS$_3\times  \text{S}^2\times \text{CY}_2$ foliated over an interval} \label{eq:classICYfoliations}
In this section we study the sub-class of local solutions that follow from section \ref{sec:classIsummery} by imposing that the symmetries of CY$_2$ are respected by the full solution. This means that the warp factors cannot depend on the directions on CY$_2$ and and we must fix $H_2=0$. As such, the only way to realise a compact internal space is if  CY$_2$ is itself compact. This restricts our considerations to
\beq
\text{CY}_2= \text{T}^4~~~~\text{or}~~~~ \text{CY}_2=\text{K3}.
\eeq
The supersymmetry condition \eqref{eq:caseIcon1} and Bianchi identities \eqref{eq:BI1} are then all completely solved for $h_8,u,h_4$  arbitrary linear functions in $\rho$. We parametrise these in general by introducing five arbitrary constants $(c_1,...,c_5)$ such that
\beq\label{eq:funccond}
h_8= c_1+ F_0 \rho,~~~~ u= c_2+ c_3\rho,~~~~ h_4= c_4+ c_5 \rho,
\eeq 
at which point the local form of a general solution in this class may be written explicitly\footnote{At least when CY$_2=\text{T}^4$. The metrics on K3 manifolds are not know explicitly, but they are known to exist by Yau's theorem.}. The NS sector of the general solution is 
\begin{align}\label{eq:NSfolationI}
ds^2&= \frac{(c_2+ c_3 \rho)}{\sqrt{c_1+ F_0 \rho}\sqrt{c_4+ c_5 \rho}}\bigg[ds^2(\text{AdS}_3)+\frac{1}{4+ \frac{c_3^2}{(c_1+ F_0 \rho)(c_4+ c_5 \rho)}}ds^2(\text{S}^2)\bigg]+\sqrt{\frac{c_4+ c_5 \rho}{c_1+ F_0 \rho}} ds^2(\text{CY}_2)\nn\\[2mm]
&+\frac{\sqrt{c_1+ F_0 \rho}\sqrt{c_4+ c_5 \rho}}{(c_2+ c_3 \rho)} d\rho^2,~~~~ e^{-\Phi}= \frac{(c_1+ F_0 \rho)^{\frac{3}{4}}\sqrt{c_3^2+ 4 (c_1+ F_0\rho)(c_4+ c_5 \rho)}}{2\sqrt{c_2+ c_3 \rho}(c_4+ c_5 \rho)^{\frac{1}{4}}},\nn\\[2mm]
H&=dB_2,~~~~B_2= \frac{1}{2}\left(2n \pi-\rho+\frac{c_3(c_2+ c_3 \rho)}{c_3^2+ 4(c_4+ c_5\rho)(c_1+ F_0\rho)}\right)\wedge\text{vol}( \text{S}^2),
\end{align}
where we have added the closed form $n\pi\text{vol}( \text{S}^2)$ to $B_2$ that parametrises large gauge transformations - so $n$ is an integer. The 10 dimensional RR fluxes follow from substituting \eqref{eq:funccond} and $H_2=0$ into  \eqref{eq:classIflux1}-\eqref{eq:classIflux3}. However, in what follows we will find it more useful to know the magnetic parts of the Page fluxes explicitly, ie $\hat f= f\wedge e^{-B_2}$, where $f$ encloses the magnetic components of the 10 dimensional RR fluxes. We find
\begin{subequations}\label{eq:RRfolationI}
\begin{align}
\hat f_0&= F_0,~~~~\hat f_2=- \frac{1}{2}(c_1+ 2n \pi F_0)\text{vol}( \text{S}^2),\label{eq:NSfolationIf1}\\[2mm]
\hat f_4&=- c_5\text{vol}(\text{CY}_2) ,~~~~\hat f_6= \frac{1}{2}(c_4+  2n \pi c_5 ) \text{vol}(\text{CY}_2)\wedge \text{vol}( \text{S}^2)\label{eq:NSfolationIf2}.
\end{align}
\end{subequations} 
Flux quantisation for Dp brane like objects requires that the Page charges, $N_{p}= \frac{1}{(2\pi)^{7-p}}\int_{\Sigma_{8-p}} \hat f_{8-p}$, are integer. This requires that one tunes
\begin{align}\label{eq;caseItunings}
&2\pi F_0= N_8,~~~~-\frac{c_5}{(2\pi)^3}\int_{\text{CY}_2}\text{vol}(\text{CY}_2)=N_4,\nn\\[2mm]
&-c_1= n_6,~~~~~~\frac{c_4}{(2\pi)^4}\int_{\text{CY}_2}\text{vol}(\text{CY}_2)=n_2,
\end{align}
where $n_i\in \mathbb{Z}$, so that we have the integer Page charges $N_8,N_4$ and
\beq
N_6= n_6 - n N_8,~~~~~ N_2= n_2- n N_4.
\eeq
Of course, as they are defined in terms of arbitrary constants, not all these integers need to be non-zero in a given solution. The holographic central charge of a generic solution in this class at leading order is then given by
\beq\label{eq:centralchargecaseI}
c_{\text{hol}}= \frac{3}{2^4\pi^6}\int_{\text{M}_7}e^{A-2\Phi}\text{vol}(\text{M}_7) = \frac{3}{4\pi^3}\int d\rho (2\pi n_2 - N_4\rho)(N_8\rho-2\pi n_6 ),
\eeq
where we have converted the formula of \cite{Couzens:2017way} to string frame. However one needs to know the range of $\rho$ to perform the final integration - which depends on how $c_i$ are tuned. For similar reasons the charge associated to $H$, which is defined on $(\rho, \text{S}^2)$ needs to be computed on a case by case basis.

A brief glance at \eqref{eq:NSfolationI} makes it clear that the generic solution in this section is not regular, though regularity can be achieved by tuning $c_i$.  A regular solution requires the AdS$_3$ warp factor to be either constant, or constant at the boundaries of the interval spanned by $\rho$. Only the former leads to a compact space in this case, and requires tuning $h_8 \propto u \propto h_4$.  We thus set
\beq\label{eq:hybridlimit}
c_2= L^2 \lambda^2c_1,~~~~c_3= L^2  \lambda^2 F_0,~~~~c_4= \lambda^4c_1,~~~~c_5=\lambda^4 F_0,
\eeq
without loss of generality. The metric then reduces to 
\beq\label{eq:TNATDhybrid}
 ds^2= L^2 ds^2(\text{AdS}_3)+ \lambda^2 ds^2(\text{CY}_2)+ \frac{1}{ L^2 }d\rho^2+\frac{L^2(c_1+ F_0 \rho)^2}{L^4 F_0^2+ 4(c_1+ F_0 \rho)^2}ds^2(\text{S}^2).
\eeq
This solution is regular: When $F_0= 0$ the warp factors are constant so this point is trivial, $\partial_{\rho}$ becomes an isometry and the metric is compact if we make it parametrise a circle. For generic $F_0$, $\rho$ is bounded from below at $\rho= -\frac{c_1}{F_0}$, where the sub-manifold spanned by $(\rho, \text{S}^2)$ vanishes as $\mathbb{R}^3$ in polar coordinates. However $\rho$ is not bounded from above, with $\rho=\infty$ at infinite proper distance, so the metric is non compact. In fact when $F_0=0$ \eqref{eq:TNATDhybrid} is the metric of the T-dual of the D1-D5 near horizon geometry, and taking $0<\rho<2\pi$ in the formula for the central charge \eqref{eq:centralchargecaseI} yields $c_{\text{hol}}= 6 N_2 N_6$, as one expects for this class. In turn, for $F_0\neq 0$ it is the non-Abelian T-dual of this system\footnote{This is perhaps made more obvious if one defines $r=c_1+ F_0 \rho$ and substitutes for $\rho$ in favour of $r$.}. These observations extend to the fluxes and dilaton as well. As such, the solution defined by \eqref{eq:hybridlimit} is somewhat novel, in that it gives a hybrid solution containing both the T and non-abelian T-duals of a known solution.

For choices of $c_i$ other than \eqref{eq:hybridlimit}, the metric in \eqref{eq:NSfolationI} will necessarily contain singular loci, making the solution non regular. However non regularity is not always a reason to worry. Indeed, there are situations in which one can trust a singularity in a supergravity solution, namely when it signals the presence of a physical object in string theory and when the radius of divergent behaviour about this object, where the supergravity approximation does not hold, can be made arbitrarily small by tuning parameters. Given the form of  \eqref{eq:NSfolationI}  we anticipate D brane and O plane sources. Thus, we will allow $\rho$ to terminate at a singular point of the space when the  solution reduces to the behaviour of these objects at this loci, ie if the leading order behaviour of the metric and  dilaton are diffeomorphic to one of the following forms
\begin{align}\label{eq:boundaries}
\text{Dp brane}&:~~~~ ds^2\sim r^{\frac{7-p}{2}}ds^2(\text{M}^{1,p})+ r^{\frac{-7+p}{2}}\bigg(dr^2 + r^2 ds^2(\text{B}^{8-p})\bigg),~~~ e^{\Phi} \sim r^{\frac{(3-p)(-7+p)}{4}},\nn\\[1mm] 
\begin{array}{c}\text{Dp smeared}\\ \text{on $\tilde{B}^s$}\end{array}&:~~~~ds^2\sim r^{\frac{7-p-s}{2}}ds^2(\text{M}^{1,p})+ r^{\frac{-7+p+s}{2}}\bigg(dr^2 +ds^2(\tilde{\text{B}^s})+ r^2 ds^2(\text{B}^{8-p-s})\bigg),~~~ e^{\Phi} \sim r^{\frac{(3-p)(-7+p+s)}{4}},\nn\\[1mm] 
\text{Op plane}&:~~~~ ds^2\sim \frac{1}{\sqrt{r}} ds^2(\text{M}^{1,p})+\sqrt{r} \bigg(dr^2 + r_0^2 ds^2(\text{B}^{8-p})\bigg),~~~ e^{\Phi} \sim r^{\frac{3-p}{4}}.
\end{align}
Here $\text{M}^{1,p}$ is some manifold that the object wraps, $\text{B}^{8-p}$ a compact base, on which one integrates to get the associated charge of this object, and $\tilde{\text{B}}^s$ is the manifold over which a brane is smeared. We have included smeared D branes but not O planes, because the former is dynamical in string theory while the latter is not. We shall also allow for coincident objects such as a Dp-brane inside the world volume of a D(p+4)-brane\footnote{Both must depend on the same radial variable for this behaviour to occur, hence the Dp-brane is smeared over the remaining world volume directions of the D(p+4)-brane.}. If $\text{M}^{1,p}$ were flat, the only magnetic flux near a Dp/Op singularity would be $f_{8-p}$ - but here $\text{M}^{1,p}$ will not be flat, so one should expect additional fluxes to be turned on at the singularity to accommodate this. 

By tuning the constants $c_i$ we are able to find a rich variety of physical boundary behaviours, namely
\beq\label{eq:boundary behaviours}
\begin{array}{c|c|c|c|c}
\text{Source} & \text{Minimal tuning} & \text{M}^{1,p} & \tilde{\text{B}}^{s} & \text{Loci}  \\[2mm]
\hline
\text{D8/O8} & c_3=0 & \text{all but}~ \rho & - & \rho= -\frac{c_1}{F_0}\\
\text{D6} & c_3\neq 0~~~c_4=b c_2,~~~ c_5= b c_3& \text{AdS}_3\times \text{CY}_2 & - & \rho= -\frac{c_2}{c_3}\\
\text{O6} & \text{generic}~c_i & \text{AdS}_3\times \text{CY}_2 & -&\rho= -\frac{c_1}{F_0} \\
\text{Smeared D4}& c_3 =0& \text{AdS}_3\times  \text{S}^2 & \text{CY}_2  & \rho= -\frac{c_4}{c_5}\\
\text{Smeared D2}_1&c_2 =b c_1,~~~ c_3= b F_0 \neq 0   & \text{AdS}_3 & \text{CY}_2& \rho= -\frac{c_2}{c_3}\\
\text{Smeared D2}_2&\text{generic}~c_i   & \text{AdS}_3 &  \text{S}^2\times \text{CY}_2& \rho= -\frac{c_4}{c_5}\\
\text{D4 in D8}&c_3=0,~~~ c_4=bc_1,~~~c_5=b F_0   & \text{D8}:  \text{all but}~ \rho& \text{D4}:~\text{CY}_2 & \rho= -\frac{c_1}{F_0}\\
\text{D2 in D6}&\text{generic}~c_i   & \text{D6}: \text{AdS}_3\times \text{CY}_2& \text{D2}:~\text{CY}_2 & \rho= -\frac{c_2}{c_3}\\
\text{O2 in O6}&c_4=b c_1,~~~ c_5= bF_0   & \text{O6}: \text{AdS}_3\times \text{CY}_2& \text{O2}:~ \text{S}^2\times\text{CY}_2 & \rho= -\frac{c_4}{c_5}\\
\text{T/NATD hybrid} &\eqref{eq:hybridlimit}     & -& -& -\\
\end{array}\nn
\eeq 
Here $b$ is an arbitrary constant, and we have included the T-dual and non-Abelian T-dual hybrid solution, which is regular, as well as smeared O2 inside an O6 plane, for completeness. Note that when the D4, D2 branes are delocalised on all directions but $\rho$, one could also interpret them as smeared O4 and O2 planes respectively. To realise a compact solution from \eqref{eq:NSfolationI}  beyond the $F_0=0$ limit of  \eqref{eq:TNATDhybrid},  we need two of these boundary behaviours to exist for the same tuning of $c_i$. There are in fact several such solutions for various tunings of $c_i$.  In summary, we see that the following behaviours can exist simultaneously
\beq\label{eq:boundary behaviours2}
\begin{array}{c|c|c}
\text{Tuning} & \text{Boundary behaviours}& \text{Loci:~}\rho=\\[2mm]
\hline
\text{generic}~c_i& \text{O6}~|~ \text{D2 in D6}~|~ \text{D2}_2 &  -\frac{c_1}{F_0}~|~-\frac{c_2}{c_3}~|~-\frac{c_4}{c_5}\\
(c_4= b c_1,c_5= b F_0,c_3\neq 0) &\text{O2 in D6}~|~\text{D2 in O6} & -\frac{c_2}{c_3}~|~-\frac{c_1}{F_0}\\
(c_2= b c_4,c_3= b c_5,c_3\neq 0) & \text{D6}~|~ \text{O6} & -\frac{c_2}{c_3}~|~-\frac{c_1}{F_0}\\
(c_2= b c_1,c_3= b F_0,c_3\neq 0) & \text{D2}_1~|~\text{D2}_2 &-\frac{c_1}{F_0}~|~-\frac{c_4}{c_5}\\
c_3=0&\text{D8/O8}~|~ \text{D4}~ & -\frac{c_1}{F_0} ~|~ -\frac{c_4}{c_5},
\end{array}
\eeq
making for a total of 7 independent compact solutions of this type\footnote{Really 14, as we can take $\text{CY}_2=\text{T}^4$ or K3 for each.}. In the interest of (relative) brevity we are going to look only at the two simplest cases explicitly, those with interval bounded between D8s/O8 and D4s, and those between D6 and an O6.\\
~\\
\textbf{Interval bounded between D8/O8s and D4s}\\
To realise the first compact example with sources, one should tune $c_3=0$. Then  one of $(c_1,c_4)$ can also be set to zero with a coordinate transformation of $\rho$. Here we take $c_4=0$. The resulting NS sector is
\begin{align}
ds^2&= \frac{c_2}{\sqrt{ c_5}\sqrt{ \rho}\sqrt{c_1+ F_0 \rho}}\bigg[ds^2(\text{AdS}_3)+ \frac{1}{4} ds^2( \text{S}^2)\bigg]+\frac {\sqrt{ c_5}\sqrt{ \rho}}{\sqrt{c_1+ F_0 \rho}}ds^2(\text{CY}_2)+ \frac{\sqrt{ c_5}\sqrt{ \rho}\sqrt{c_1+ F_0 \rho}}{c_2}d\rho^2,\nn\\[2mm]
e^{-\Phi}&= \frac{c_5^{\frac{1}{4}}\rho^{\frac{1}{4}}(c_1+ F_0 \rho)^{\frac{5}{4}}}{\sqrt{c_2}},~~~~B_2= \left(n \pi-\frac{\rho}{2}\right)\wedge\text{vol}( \text{S}^2),
\end{align}
and the magnetic RR Page fluxes are given by \eqref{eq:NSfolationIf1}-\eqref{eq:NSfolationIf2} with $c_3=c_4=0$. It should not be hard to see that close to $\rho=-\frac{c_1}{F_0}=\frac{2\pi n_6}{N_8}$ the metric and dilaton are consistent with D8 and O8 behaviour, while at $\rho=0$ they signal  D4 branes wrapped on AdS$_3\times \text{S}^3$ and smeared on CY$_2$. This bounds the interval as $0<\rho< \frac{2\pi n_6}{N_8}$, assuming $\frac{2\pi n_6}{N_8}>0$.  This solution is also under parametric control, with the supergravity approximation holding for $c_5\sim N_4>>1$,  with the radius of divergent behaviour about the poles scaling inversely with this parameter. Given the flux quantisation conditions \eqref{eq;caseItunings}, the Page charges are then
\beq
N_8,~~~~N_4,~~~~N_6= n_6 - n N_8,~~~~N_2= - n N_4,~~~~N_5= \frac{n_6}{N_8},
\eeq
where $N_5= \frac{1}{(2\pi)^2}\int_{(\rho, \text{S}^2)}dB_2$ is the charge associated to NS5 branes. We now turn our attention to the large gauge transformations parameterised by $n$ in the definition of the NS 2-form. As $(\rho, \text{S}^2)$ defines a  cycle at each of the singular loci, we should impose that $b= -\frac{1}{(2\pi)^2}\int_{ \text{S}^2}B_2$ should be an integer at these points. This may be achieved by constraining
\beq
b=\frac{\rho}{2\pi}- n ~~~\text{s.t}~~~ 0\leq b <1,
\eeq
which implies that $\rho$ is partitioned into segments of length $2\pi$. At $\rho=0$, $b=n$ so we can take $n=0$  fixing $b=0$. One should then perform a large gauge transformation sending $n\to n+1$ each time $\rho$ increases by $2\pi$. At $\rho=\frac{2\pi n_6}{N_8}$ one has $b=N_5+m$ for $m$ the number of gauge transformations required to traverse the interval. Finally we can integrate the expression for the holographic central charge at leading order \eqref{eq:centralchargecaseI}, which yields 
\beq\label{eq:confusing}
c_{\text{hol}}= n_6 N_4 N_5^2.
\eeq
At first sight this may appear confusing as the (left) central charge of small $\mathcal{N}=(4,0)$ CFTs should be related to the level of the affine SU(2) algebra as $c=6k$. Here \eqref{eq:confusing} contains no factor of 6, but one should recall that  $c_{\text{hol}}$ is the central charge in the supergravity limit $N_5>>1$, which only gives the leading order contribution to $c$, neglecting sub-leading terms in $N_5$. We believe that one will recover $c=6 k$ if one includes the 1-loop correction to $c_{\text{hol}}$ - however to our knowledge this corrections is not yet known for massive IIA, so one cannot yet check this explicitly. Adding support to this claim is \cite{LMNR2}, where in section 4 the central charges of several concrete CFTs and geometries that are locally of the form \eqref{eq:RRfolationI} are compared. The CFTs obey $c=6k$, but to leading order in some parameter(s), where it makes sense to compare to supergravity, the factor of 6 is lost in many cases - non the less $c=c_{\text{hol}}$ in these limits.\\
~\\
\textbf{Interval bounded between D6s and an O6}\\
One can get the second compact solution by tuning $c_2= b c_4,c_3= b c_5$. One can then set $c_4=0$ without loss of generality with a diffeomorphism. This results in the following NS sector
\begin{align}
ds^2&= \frac{b\sqrt{ c_5}\sqrt{ \rho}}{\sqrt{c_1+ F_0 \rho}}\bigg[ds^2(\text{AdS}_3)+ \frac{\rho(c_1+ F_0 \rho) }{b^2 c_5+4\rho(c_1+ F_0 \rho)} ds^2( \text{S}^2)\bigg]+\frac {\sqrt{ c_5}\sqrt{ \rho}}{\sqrt{c_1+ F_0 \rho}}ds^2(\text{CY}_2)+ \frac{\sqrt{c_1+ F_0 \rho}}{b\sqrt{ c_5}\sqrt{ \rho}}d\rho^2,\nn\\[2mm]
e^{-\Phi}&= \frac{\sqrt{c_1+ F_0\rho}\sqrt{b^2 c_5 +4\rho(c_1+ F_0\rho)}}{2 \sqrt{b} c_5^{\frac{1}{4}}\rho^{\frac{3}{4}}},~B= \left(n \pi-\frac{2\rho^2(c_1+ F_0 \rho)}{b^2 c_5 +4\rho(c_1+ F_0\rho)}\right)\wedge\text{vol}( \text{S}^2),
\end{align}
where again $n$ is an integer parametrising large gauge transformations and the magnetic RR Page fluxes are given by \eqref{eq:NSfolationIf1}-\eqref{eq:NSfolationIf2} with $c_2= c_4=0,c_3= b c_5$. This time one can show that the behaviour close to $\rho=-\frac{c_1}{F_0}=\frac{2\pi n_6}{N_8}$ corresponds to an O6 plane wrapped on AdS$_3\times$ CY$_2$, while that of  $\rho=0$ is a D6 brane - bounding the interval  as $\rho\in(0,\frac{2\pi n_6}{N_8})$. The supergravity approximation is valid this time for $F_0>>c_1\sim n_6>>0$. The Page charges are
\beq
N_8,~~~~N_4,~~~~N_6= n_6 - n N_8,~~~~N_2= - n N_4,~~~~N_5= 0,
\eeq
with the major difference with respect to the previous example being that the NS charge $N_5=0$. We should again constrain  
\beq
b= -\frac{1}{(2\pi)^2}\int_{ \text{S}^2}B_2~~~\text{s.t}~~~0<b<1.
\eeq
The form of $B_2$ means that the $\rho$ dependence vanishes at the boundaries and reaches an extrema in between at $\rho=\rho_0$, which depends non trivially on the charges. As at $\rho=0$ one sets $n=0$, then as one traverses the interval $0<\rho<\rho_0$  successive large gauge transformations are needed to keep $0<b<1$. But once one crosses $\rho=\rho_0$ one needs to start undoing the gauge transformations to keep $b$ bounded until $n=0$, once more at $\rho=\frac{2\pi n_6}{N_8}$. From this we conclude that the charge of both the D6s and the O6 is equal to $n_6$, which gives a problem. Weak curvature requires $n_6>>0$, but the charge of the O6 is fixed to be $\pm 4$. As such, the solution is strongly curved everywhere.  Nonetheless, for the sake of comparison we  compute the holographic central charge from \eqref{eq:centralchargecaseI}, and find
\beq
c_{\text{hol}}=\frac{n_6^3 N_4}{N_8^2},
\eeq
~\\
\textbf{Summary of this section}\\
To summarise, in this section \ref{eq:classICYfoliations} we have studied the local solutions of class I that  respect the symmetry of $\text{CY}_2$ - they are foliations of AdS$_3\times  \text{S}^2 \times \text{CY}_2$ over an interval. We have found that the general solution can be given explicitly and depends on parameters $(c_1,...c_5,F_0)$, with all necessary conditions solved. Different behaviours can be achieved by tuning these parameters, and we have found an array of physical boundary behaviours for the interval. We have established that there are 8 independent compact solutions: The T-dual of the D1-D5 near horizon, which is regular, and the 7 independent combinations one can form from \eqref{eq:boundary behaviours2} with various physical singularities. We have chosen two of these solutions for a more detailed study, where the interval is bounded between either D8/O8s and D4 or D6s and an O6. We have shown that only the former has a good interpretation in supergravity, with the latter requiring higher curvature corrections.\\
\\~
Before moving on let us first stress that the general solution of this section is only a local one. What this really means is that every coordinate patch of a global solution can be expressed in the form of  \eqref{eq:NSfolationI} and \eqref{eq:NSfolationIf1}-\eqref{eq:NSfolationIf2}, but the specific values of $(F_0,c_1,...,c_5)$   in each of these patches may differ in principle. This fact was exploited in \cite{Apruzzi:2013yva} to construct infinite classes of globally compact AdS$_7$ solutions in massive IIA, by glueing together various non compact solutions with defect branes. We shall return to this issue in section \ref{sec:calibrations}, where we will establish that this is also possible for the local solutions of this section. We delay a detailed analysis of such solutions until \cite{LMNR2,LMNR3}.\\
\\~
In the next section we shall begin our analysis of the second class of solutions we consider in this paper, namely those containing a family of Kahler four-manifolds. 
\section{Class II: Kahler four-manifold case}\label{sec: Class II}
In this section we study the second class of solutions following from the necessary conditions of section \ref{sec:geometricsusy} for $\sin\beta \neq 0$. We find that the solutions decompose as a warped product of AdS$_3\times  \text{S}^2 \times \hat{\text{M}}_4\times \mathbb{R}$ where $\hat{\text{M}}_4$ is a family of Kahler manifolds with metrics that depend on the interval.

In section \ref{sec:classIIsummery} we summarise class II and discuss some of its general features, deferring its derivation to section \ref{sec:classIIderivation}.  In section \ref{eq:D3curves} we T-dualise class II along the interval and arrive at a generalisation of \cite{Couzens:2017way} with non trivial 3-form flux. And finally in section \ref{eq:classIICYfoliations} we expand up section \ref{eq:classICYfoliations} and present further local solutions that are foliations of AdS$_3\times  \text{S}^2\times$CY$_2$ over an interval. 

\subsection{Summary of class II}\label{sec:classIIsummery}
The solutions in  class II have the following NS sector
\begin{align}\label{eq:classIINS}
ds^2&= \frac{u}{\sqrt{h w^2- v^2}}\bigg[ds^2(\text{AdS}_3)+\frac{h w^2- v^2}{4 (h w^2- v^2)+ (u')^2} ds^2( \text{S}^2)\bigg] +\frac{\sqrt{h w^2- v^2}}{u}\bigg[ \frac{u}{h w}ds^2(\hat{\text{M}}_4)+ d\rho^2\bigg],\nn\\[2mm]
H&=\frac{1}{2}d\left(-\rho+ \frac{ u u'}{4 (h w^2- v^2)+ (u')^2}\right)\wedge \text{vol}( \text{S}^2)+ d\left(\frac{v}{w h}\hat J\right),\nn\\[2mm]
e^{-\Phi}&= \frac{w h^{\frac{1}{2}}\sqrt{4 (h w^2- v^2)+ (u')^2}}{2 \sqrt{u} (h w^2- v^2)^{\frac{1}{4}}}.
\end{align}
Here $\hat{\text{M}}_4$ is a family of Kahler manifolds parameterised by $\rho$, 
 with an integrable complex structure that is $\rho$ independent. 
  $\hat{J}$ is a two-form defined on the Kahler four-manifold (the details are given below). 
 The functions $u,v,w$ depend on $\rho$ only, while $h$ has support in $(\rho,\hat{\text{M}}_4)$. In fact $w$ is actually redundant, as it can be absorbed into $h$ and $\hat{\text{M}}_4$. We keep it for convenience as it simplifies the derivation of the classes  in sections \ref{eq:D3curves} and \ref{eq:classIICYfoliations}.  
Supersymmetry is ensured by the following differential conditions
\beq\label{eq:caseIIcon1}
u''=0,~~~~\partial_{\rho}\left(\frac{ \hat{g}^{\frac{1}{2}}}{h}\right)=0,~~~~i\partial \overline{\partial}\log h = \hat{\mathfrak{R}},
\eeq
for $\hat{\mathfrak{R}}$ the Ricci form and $\hat{g}$ the determinant of the metric on $\hat{\text{M}}_4$. $\partial, \overline{\partial}$ are Dolbeault operators expressed in terms of complex coordinates on $\hat{\text{M}}_4$ such that $d_4=\partial+ \overline{\partial}$.  The 10 dimensional RR fluxes of this class take the form
\begin{align}
F_0 &= v',\nn\\[2mm]
F_2&=-\frac{w^2}{u} d\rho\wedge \hat\star_4(d_4h\wedge \hat J)-\partial_{\rho}(w\hat J)+\frac{v v'}{h w}\hat J- \frac{1}{2}\left(v- \frac{v'u u'}{4(h w^2- v^2)+ (u')^2}\right)\text{vol}( \text{S}^2),\nn\\[2mm]
F_4&=\frac{1}{2}\text{vol}(\text{AdS}_3)\wedge\bigg(d\left(\frac{v u u'}{h w^2- v^2}\right)+ 4 v d\rho\bigg)+\frac{v}{2h}\left(\frac{ v v'}{h w^2}-\partial_{\rho}\log(v^{-1} h w^2)\right)\hat J\wedge \hat J\nn\\[2mm]
&-\frac{v w}{ u} d\rho \wedge\hat\star_4d\log h+\frac{1}{2}\bigg(\frac{u u'}{4(h w^2-v^2)+ (u')^2}F_2+ \frac{h w^2-v^2}{h w}\hat J\bigg)\wedge \text{vol}( \text{S}^2),
\end{align}
where again $F_6=-\star_{10}F_4$, $F_8=\star_{10}F_2$. The Bianchi identities are then solved away from localised sources when
\beq\label{eq:caseIIcon2}
v''=0,~~~~2i\partial\overline{\partial}h= \partial_{\rho}^2(w\hat J).
\eeq
The conditions \eqref{eq:caseIIcon1} and \eqref{eq:caseIIcon2} are necessary and sufficient for a solution to exist in the absence of sources. When these exist one should also check the source corrected Bianchi and calibration conditions at their loci.

To better understand this second class of solutions one can consider the limit $u=w=1$ and $v=0$ with $\partial_{\rho}$ an isometry. The result coincides with the Hopf fibre T-dual of the class of solutions found in \cite{Couzens:2017way}. These solutions are of the form $\text{AdS}_3\times \text{S}^3\times B$, for $B$ the base of an elliptically fibered Calabi-Yau 3-fold. They are characterised by varying axio-dilaton with D3 branes wrapped on a curve within $B$, but have no 3-form flux. If we instead consider a similar limit with  $v=$ constant rather than zero, we find a generalisation of this class with non trivial 3-form flux, as we shall  demonstrate in  section \ref{eq:D3curves}. In addition to containing the T-dual of this IIB class, class II also contains its non-Abelian T-dual, which one can realise by fixing $w\propto u \propto v$ and taking $J$ and $\hat{\text{M}}_4$ to be $\rho$ independent. This in fact gives another hybrid solution similar to \eqref{eq:TNATDhybrid}, that realises the T-dual of section \ref{eq:D3curves} when $F_0=0$ and the non-Abelian T-dual for generic $F_0$.\\
~\\
In the next section we show how class II is derived from the necessary and sufficient conditions for supersymmetry found in section \ref{sec:su2r}.

\subsection{Derivation of class II}\label{sec:classIIderivation}
For class II we assume $\sin\beta\neq 0$, and as such we are free to  divide by $\sin\beta$ which enables us to put \eqref{eq:gensusyCOND1}-\eqref{eq:gensusyCOND8}  in the form
\begin{align}
&d(e^{A-\Phi}\sin\alpha\cos\beta)\wedge V=d(e^{3A-\Phi}\sin\alpha \sin\beta)-2\mu e^{2A-\Phi} \cos\alpha \sin\beta V=0,\label{eq:susygen1}\\[2mm]
&2e^{C}+ e^A \sin\alpha=d(\frac{1}{\sin\beta}\Omega)= d(\frac{e^{-2A}}{\sin^2\beta} J)\wedge V=0,\label{eq:susygen2}\\[2mm]
&e^{2C}H_1 =- \frac{1}{2\mu} e^A V+\frac{1}{4} d(e^{2A} \sin\alpha \cos\alpha),~~~~ H_3 = d(\frac{\cos\beta}{\sin \beta}J). \label{eq:susygen3}
\end{align}
We can solve \eqref{eq:susygen1} in general by introducing two functions $ u(\rho),v(\rho)$ such that
\beq
e^{3A-\Phi} \sin\alpha \sin\beta= u,~~~ 2 e^{A-\Phi} \cos\alpha\sin\beta= u',~~~~ e^{A-\Phi} \sin\alpha\cos\beta= v.
\eeq
In contrast to case I the conditions supersymmetry imposes on $(J,\Omega)$ do not imply that ${\hat M}_4$  is conformally Calabi-Yau in general. We will see instead that it must be a family of $\rho$ dependent Kahler four-manifolds. Taking \cite{Gauntlett:2004zh} as a guide it is useful to introduce the rescaled forms and metric
\beq
J= \frac{\sin\beta}{\sqrt{h}}\hat J,~~~~\Omega= \frac{\sin\beta}{\sqrt{h}}\hat \Omega,~~~~ds^2(\hat{\text{M}}_4)= \frac{\sin\beta}{\sqrt{h}}ds^2(\hat{\text{M}}_4),
\eeq 
where we have also introduced
\beq
\frac{1}{h}= e^{4A} \sin^2\beta w^2u^{-2}, 
\eeq
with $h$ a function of $\rho$ and the coordinates on $\hat{\text{M}}_4$. Here $w=w(\rho)$ is an arbitrary function that is actually redundant, as it can be absorbed into the definition of $h$ and $\hat{\text{M}}_4$, but extracting it now simplifies later exposition. Expanding $d= d_4+ d\rho\wedge \partial_{\rho} $  as before \eqref{eq:susygen2} implies the following conditions
\begin{subequations}
\begin{align}
d_4\hat J&=0,\label{eq:torsion1}\\[2mm]
d_4\hat \Omega&=\frac{1}{2}d_4\log h\wedge \hat\Omega,\label{eq:torsion2}\\[2mm]
\partial_{\rho} \hat \Omega&=\frac{1}{2}\partial_{\rho}\log h\hat \Omega \label{eq:torsion4}.
\end{align}
\end{subequations}
The first two conditions \eqref{eq:torsion1}-\eqref{eq:torsion2} imply that $ds^2(\hat{\text{M}}_4)$ is a family of Kahler manifolds parameterised by $\rho$, with an associated complex structure that is $\rho$ independent. Since $\hat{\text{M}}_4$ is Kahler \eqref{eq:torsion2} can be expressed as
\beq\label{eq:hatP}
d_4\hat \Omega= i \hat P\wedge\hat\Omega,~~~~\hat P=-\frac{1}{2}d_4\log h\lrcorner \hat J, ~~~~ d_4\hat P= \hat{\mathfrak{R}},
\eeq
where $\hat{\mathfrak{R}}$ is the Ricci form on  $ds^2(\hat{\text{M}}_4)$, with components  $\hat{\mathfrak{R}}_{ij}= \frac{1}{2}\hat {R}_{ijkl}\hat{J}^{kl}$, for $\hat {R}_{ijkl}$ the Riemann curvature tensor on $ds^2(\hat{\text{M}}_4)$ computed at constant $\rho$.  The condition \eqref{eq:torsion4} then just serves to constrain the $\rho$ dependence of the Kahler metric such that its determinant $\hat g$ satisfies
\beq
\partial_{\rho}\left(\frac{ \hat{g}^{\frac{1}{2}}}{h}\right)=0.
\eeq
We now turn our attention to the paring conditions \eqref{eq:paring}. Although it is not possible to explicitly take the Hodge dual of every term in \eqref{eq:flux1}-\eqref{eq:flux4}, it is still possible to solve \eqref{eq:paring} explicitly by making use of \eqref{eq:torsion1}-\eqref{eq:torsion4}, \eqref{eq:hodgedualofforms}, and the following identities involving an arbitrary 1-form in 5 dimensions $U= U_4+ u_0 V$:
\begin{align}\label{eq:starrules}
&j_1\wedge \star_5( U\wedge j_2)=-j_2\wedge \star_5( U\wedge j_1)= U\wedge V \wedge j_3~~~~\text{and cyclic in 123} \\[2mm]
&j_1\wedge \star_5( U\wedge j_1)=j_2\wedge \star_5( U\wedge j_2)=j_3\wedge \star_5( U\wedge j_3)= \star_5 U_4+ u_0 j_3\wedge j_3,
\end{align}
where $J=j_3,~\Omega=j_1+ i j_2$. After a lengthy calculation we find that \eqref{eq:paring} imposes simply 
\beq
u''=0.
\eeq
Having now dealt with the geometric supersymmetry constraints we turn our attention to Bianchi identities of the RR fluxes. In general the RR fluxes are rather involved, but as is often the case, the Page fluxes $\hat f= e^{-B}\wedge f$, where $dB=H$, are rather more simple, so let us first study these. The conditions defining $H_1,H_3$ in \eqref{eq:susygen3} can  be locally integrated with ease giving rise to the NS potential
\beq
B=\frac{1}{2}\left(-\rho+ \frac{ u u'}{4 (w^2h - v^2)+ (u')^2}\right)\wedge \text{vol}( \text{S}^2)+ \frac{v}{w h}\hat J,
\eeq
in terms of which the Page fluxes take the following form\footnote{We have significantly simplified $\hat f_2, \hat f_4$ by making use of \eqref{eq:starrules} and
\beq
\partial_{\rho} \hat J=\frac{1}{2}\partial_{\rho}\log h\hat J+ H_2,~~~~H_2+ \hat\star_4 H_2=0,
\eeq
where $J\wedge H_2=\Omega\wedge H_2=0$,
which follows  from \eqref{eq:torsion4} and the allowed torsion classes of SU(2)-structures in 5 dimensions \cite{Torsion}
.}
\begin{subequations}
\begin{align}
\hat f_0&=f_0= v',\\[2mm]
\hat f_2&=-\frac{w^2}{u} d\rho\wedge \hat\star_4(d_4h\wedge \hat J)-\partial_{\rho}(w\hat J)+\frac{1}{2}(\rho v'-v) \text{vol}( \text{S}^2),\\[2mm]
\hat f_4&=\frac{v'}{2h}\hat J\wedge \hat J+\frac{1}{2}\bigg(\rho\hat f_2+ w  \hat J\bigg)\wedge \text{vol}( \text{S}^2),\\[2mm]
\hat f_6&=\bigg(\frac{\rho}{2}\hat f_4- \frac{v}{4 h}\hat J \wedge \hat J\bigg)\wedge\text{vol}( \text{S}^2).
\end{align}
\end{subequations}
Away from localised sources the Bianchi identities of the RR fluxes hold if and only if the Page fluxes are closed. Imposing this yields the conditions
\begin{subequations}
\begin{align}
&v''=0,\label{eq:BIcaseII1}\\[2mm]
&\frac{w^2}{u} d_4 \hat\star_4(d_4 h\wedge \hat J)= \partial_{\rho}^2(w\hat J),\label{eq:BIcaseII2}
\end{align}
\end{subequations}
that follow from the parts of $\hat f_0,\hat f_2$ that are orthogonal to $\text{vol}( \text{S}^2)$ - closure of the rest is implied by these and supersymmetry. We can make further progress by introducing complex coordinates  $z_1,z_2$ on $\hat{\text{M}}_4$ and  Dolbeault operators $\partial= dz^i\partial_{z_i} $, $\overline{\partial}= d\overline{z}^i\partial_{\overline{z}_i} $ in terms of which we can expand $d_4= \partial+ \overline{\partial}$. We then have 
\beq
\hat\star_4(d_4 g\wedge \hat J)=d_4\log g\lrcorner \hat J=  -i (\partial- \overline{\partial})g,
\eeq
for $g$ an arbitrary function. This can be used to simplify some of the necessary conditions, allowing us to present the class in the form given in section \ref{sec:classIIsummery}.\\
~\\
In the next section we will derive a class of solutions in IIB that generalise the solutions in  \cite{Couzens:2017way} to include non trivial 3-form flux.
\subsection{Generalisation of the F-theory solutions in  \cite{Couzens:2017way} with non trivial 3-form flux}\label{eq:D3curves}

In this section we derive a generalisation of a class of solutions in IIB found in \cite{Couzens:2017way}. These are characterised by varying axio-dilaton with D3-branes wrapped on complex curves within the base of an elliptically fibered CY$_3$, and vanishing 3-form flux. Our generalisation will include a non-trivial 3-form flux.

We begin with class II of section \ref{sec:classIIsummery} and impose that $\partial_{\rho}$ is an isometry. This can be achieved without loss of generality by fixing
\beq
u=L^4,~~~~ w=L^4\lambda^2~~~~~ v= c L^2,
\eeq
with the Kahler manifold and structure assumed to be $\rho$ independent.
We also rescale $h$ for convenience as
\beq
h \to \frac{h}{L^4 \lambda^4},
\eeq
for $(L,\lambda,c)$ all constant. The NS sector then becomes
\begin{align}\label{eq:IIA2formD3curveNS}
ds^2&= \frac{L^2}{\sqrt{h-c^2}}\bigg[ds^2(\text{AdS}_3)+ \frac{1}{4}ds^2( \text{S}^2)\bigg]+\frac{\sqrt{h-c^2}}{L^2}d\rho^2+ L^2\lambda^2\frac{\sqrt{h-c^2}}{h} ds^2(\hat{\text{M}_4}),\nn\\[2mm]
B&=-\frac{1}{2} d\rho\wedge \eta+ c L^2\lambda^2  h^{-1}\hat J,~~~~ e^{-\Phi}=  L\sqrt{h} (h-c^2)^{\frac{1}{4}},
\end{align}
as before, $d\eta= -\text{vol}( \text{S}^2)$ and $dB=H$ - we have chosen a
 gauge for $B$ that makes the  $\partial_{\rho}$ isometry explicit . The RR sector becomes
\begin{align}
F_2&= i d\rho\wedge(\partial-\overline{\partial})h- \frac{L^2 c}{2} \text{vol}( \text{S}^2),\\[2mm]
F_4&=2 c L^2 \text{vol}(\text{AdS}_3)\wedge d\rho+ cL^2\lambda^2 (\hat\star_4 d\log h)\wedge d\rho+\frac{L^4 \lambda^2}{2} \frac{h-c^2}{h} \hat J\wedge \text{vol}( \text{S}^2),
\end{align}
with $F_0=0$. The first thing we note is that the only Bianchi identity that is not solved automatically is that of the RR 2-form, due to the first term. Whenever this is satisfied away from localised sources there exists a local function $C_0$ with support on $\hat{\text{M}}_4$ such that
\beq\label{eq:C0}
dC_0=i(\partial-\overline{\partial})h,
\eeq
which holds precisely when the following complex function is holomorphic 
\beq
\sigma= C_0+ i h,
\eeq
ie $\overline{\partial} \sigma=0$ implies \eqref{eq:C0} and vice-versa. This is already very reminiscent of \cite{Couzens:2017way}. Indeed if we fix $c=0$ we reproduce the result of T-dualising that class on the Hopf fibre of the $\text{S}^3$, with $\tau=\sigma$ the modular parameter of IIB. Generic $c\neq 0$ is a parametric deformation of this that obeys the same supersymmetry constraint, namely that
\beq
i\partial \overline{\partial}\log h=\frac{1}{2}d\left(\frac{d C_0}{h}\right) = \hat{\mathfrak{R}},
\eeq
which reproduces the geometric condition the base of the elliptically fibered CY$_3$ manifolds of \cite{Couzens:2017way} must obey.
After performing the T-duality on $\partial_{\rho}$, under the assumption it has period $2\pi$, the IIB string frame solution becomes
\begin{align}\label{sol:IIB}
ds^2&= \frac{L^2}{\sqrt{h-c^2}}\bigg[ds^2(\text{AdS}_3)+ds^2( \text{S}^3)\bigg]+ L^2\lambda^2\frac{\sqrt{h-c^2}}{h} ds^2(\hat{\text{M}_4}),\nn\\[2mm]
\hat B&= c \lambda^2 L^2 h^{-1}\hat J,~~~~ e^{-\hat\Phi}= \sqrt{h} \sqrt{h-c^2},\nn\\[2mm]
F_1&= dC_0,~~~~F_3=-2 cL^2 \text{vol}(\text{AdS}_3) -cL^2\lambda^2  \hat\star_4 d\log h+2 cL^2 \lambda^2 \text{vol}(\text{S}^3),\nn\\[2mm]
F_5&=-2 L^4 \lambda^2 \frac{h-c^2}{ h}(1+ \star_{10})  \hat J\wedge \text{vol}(\text{S}^3),
\end{align}
with $\hat\Phi$ and $\hat B$ the dilaton and NS 2-form potential in IIB, and where in particular $c=0$ fixes $H=F_3=0$ as in  \cite{Couzens:2017way}. Although we choose to write this solution with an S$^3$, this is meant just locally. One could equally well replace the $\text{S}^3$ with a Lens space without breaking any further supersymmetry, as is done in \cite{Couzens:2017way} by sending S$^3\to \text{S}^3/\mathbb{Z}_k$.\\
~\\
In summary, we find a parametric deformation of the solutions of \cite{Couzens:2017way} with 3-form flux turned on that still preserves $\mathcal{N}=(4,0)$ supersymmetries.  Converting to  Einstein frame, in which the  SL$(2,\mathbb{R})$ invariance of IIB is manifest, and replacing the S$^3$ with a Lens space, we arrive at the IIB  solution 
\begin{align}
ds^2_E&= L^2\bigg[\frac{h^{\frac{1}{4}}}{(h-c^2)^{\frac{1}{4}}}\bigg(ds^2(\text{AdS}_3)+ds^2( \text{S}^3/\mathbb{Z}_k)\bigg)+  \lambda^2\frac{(h-c^2)^{\frac{3}{4}}}{h^{\frac{3}{4}}} ds^2(\hat{\text{M}}_4)\bigg],\\[2mm]
 \tau &= C_0+ i \sqrt{h} \sqrt{h-c^2},~~~~ B= c \lambda^2 L^2 h^{-1}\hat J,~~~~F_5=-2 L^4 \lambda^2 \frac{h-c^2}{ h}(1+ \star_{10})  \hat J\wedge \text{vol}(\text{S}^3/\mathbb{Z}_k)\nn\\[2mm]
F_3&=-2c L^2 \text{vol}(\text{AdS}_3) -cL^2\lambda^2  \hat\star_4 d\log h+2 c L^2 \lambda^2\text{vol}(\text{S}^3/\mathbb{Z}_k)\nn.
\end{align}
This coincides locally with the class in  \cite{Couzens:2017way} when $c=0$, so that $\tau=C_0+ i h$ \footnote{The specific map is $L\to e^A,\lambda\to m_B^{-1}$.}, with AdS radius $m=1$. The complex 3-form is defined as $G= i (\text{Im}\tau)^{-1}(\tau dB-F_3)$.  Supersymmetry and the Bianchi identities away from sources, simply require
\beq
\overline{\partial} (C_0+ i h)=0,~~~~~\frac{1}{2} d\left(\frac{d C_0}{h}\right) = \hat{\mathfrak{R}}.
\eeq
Thus, as solutions were argued to exist when $c=0$, further solutions must exist for $c\neq 0$ (at least formally) as the necessary conditions for their existence are $c$ independent. A difference is that now the physical region of $\hat{\text{M}}_4$ when embedded into 10 dimensions is the portion for which $h\geq c^2$ is satisfied, with the lower bound a singular loci in the full space. The warp factors appear consistent with D5 branes wrapped on  $\text{S}^3$ at this loci\footnote{We rule out a 3-cycle in $\hat{\text{M}}_4$ because the cycle on which the D5s are wrapped should be calibrated. This ultimately means that the DBI action of the 5 brane should be equal to the pull back of some combination of the structure forms wedged with themselves, $\text{vol}(\text{AdS}_3)$ and $\hat B$. But since the structure group of $\hat{\text{M}}_4$ is SU(2) we only have two forms at our disposal  - thus any supersymmetric brane, D5 or otherwise, must wrap the $\text{S}^3$.}, however confirming this seems dependent on the specifics of the Kahler Manifold. Let us stress that, similar to the class of solutions in section \ref{sec:D5CY2}, this is not the most general  class of this type. Rather this is a specific SL(2,$\mathbb{R}$) duality frame of this most general solution - see the discussion below \eqref{eq:refto4point3}.

It would be interesting to study the solutions in this class, but as with the $c=0$ limit, the permissible metrics on $\hat{\text{M}}_4$ are the possible bases of an elliptically fibered CY$_3$, which are not explicitly known\footnote{Strictly speaking this base could be CY$_2$, in which case one could take $\text{T}^4$ as an explicit metric. However constancy requires that $h$ is constant whenever the Kahler manifold is Ricci flat.}. In particular, it would be interesting to find their explicit F-theory realisation  \cite{Lawrie:2016axq,Couzens:2017way}.

\subsection{Further local AdS$_3\times  \text{S}^2\times \text{CY}_2$ foliations}\label{eq:classIICYfoliations}
In this section we shall explore the solutions contained in class II that are foliations of AdS$_3\times  \text{S}^2\times \text{CY}_2$ over the interval spanned by $\rho$, similar to those found in section \ref{eq:classICYfoliations} - here we will be more brief. Such solutions should respect the isometries of  CY$_2$, which means the warp factors must be independent of the directions on CY$_2$. Again  CY$_2$ should be compact, which reduces our considerations to CY$_2= \text{T}^4$ or CY$_2=$ K3. The supersymmetry conditions and Bianchi identities of the fluxes (away from the loci of sources) then just impose that $(v,u,w)$ are linear functions, that we choose as 
\beq
v= c_1+ F_0\rho,~~~~u=c_2+ c_3\rho,~~~~~w=c_4+ c_5\rho,~~~h=1
\eeq
where $c_i$ are all constants. The NS sector is then
\begin{align}
ds^2&=\frac{c_2+ c_3\rho}{\sqrt{ (c_2+ c_3\rho)^2- ( c_1+ F_0\rho)^2}}\bigg[ds^2(\text{AdS}_3)+\frac{ 1 }{4 +\frac{ c_3^2}{(c_2+ c_3\rho)^2- ( c_1+ F_0\rho)^2}} ds^2( \text{S}^2)\bigg]\nn\\[2mm]
 &+\frac{\sqrt{ (c_2+ c_3\rho)^2- ( c_1+ F_0\rho)^2}}{c_2+ c_3\rho}\bigg[ \frac{c_2+ c_3\rho}{c_4+ c_5\rho}ds^2(\text{CY}_2)+ d\rho^2\bigg],\nn\\[2mm]
e^{-\Phi}&=\frac{((c_4+ c_5\rho)^2-(c_1+ F_0\rho)^2)^{\frac{3}{4}}\sqrt{4+ \frac{c_3^2}{(c_4+ c_5\rho)^2-(c_1+ F_0\rho)^2}}\sqrt{1+\frac{(c_1+ F_0 \rho)^2}{(c_4+ c_5\rho)^2-(c_1+ F_0\rho)^2}}}{2\sqrt{c_2+c_3\rho}},\nn\\[2mm]
B&= n \pi \text{vol}( \text{S}^2)-\frac{c_1+ F_0\rho}{c_4+ c_5\rho}\hat{J}- \frac{1}{2}\left(\rho-\frac{c_3(c_2+ c_3 \rho)}{4((c_4+c_5\rho)^2-
(c_1+ F_0\rho)^2)+ c_3^2}\right)\text{vol}( \text{S}^2),
\end{align}
where $n$ is an integer with which we parametrise potential large gauge transformations of the NS 2-form $B$. The magnetic Page fluxes, $\hat f=e^{-B}\wedge f$, for $f$ the magnetic components of the 10 dimensional RR fluxes, are
\begin{align}
\hat f_0&=F_0,~~~~\hat f_2= -c_5\hat {J}-\frac{1}{2}(c_1 + 2\pi n F_0) \text{vol}( \text{S}^2),\\[2mm]
\hat f_4&= F_0\hat {J}\wedge \hat {J}+\frac{1}{2}(c_4+ 2 \pi n c_5) \hat{J}\wedge\text{vol}( \text{S}^2),~~~~\hat f_6=-\frac{1}{2}(c_1+ 2\pi n F_0)\hat {J}\wedge \hat{J}\wedge\text{vol}( \text{S}^2)
\end{align}
with $F_0$ non trivial generically, and where $\hat J$ is the Kahler form of CY$_2$ (so $d\hat J=0$). 


As was the case in section \eqref{eq:classICYfoliations}, the only way to have a regular solution is if the AdS$_3$ warp factor is constant. We can achieve this by fixing
\beq\label{eq:TNATdualhybrid2}
u=L^4,~~~~ w=L^4\lambda^2~~~~~ v= c L^2,~~~~h = \frac{1}{L^4 \lambda^4}
\eeq
without loss of generality. The resulting metric takes the form
\beq\label{eq:hybrid2}
ds^2= \frac{L^2}{\sqrt{1-c^2}}\bigg[ds^2(\text{AdS}_3)+ \frac{1}{4}ds^2( \text{S}^2)\bigg]+\frac{\sqrt{1-c^2}}{L^2}d\rho^2+ L^2\lambda^2\sqrt{1-c^2} ds^2(\text{CY}_2).
\eeq
When $F_0=0$ this reproduces the metric of \eqref{eq:IIA2formD3curveNS}, in the limit $h=1$ and $\hat{\text{M}}_4=\text{CY}_2$, which is the T-dual of the same limit of the IIB solution derived in the previous section - this solution is compact when $\text{CY}_2$ and $\partial_{\rho}$ are assumed to be. For generic values of $F_0\neq 0$ the solution is the non abelian T-dual of this IIB solution, where the interval spanned by $\rho$ becomes semi infinite, with a regular zero at $\rho=-\frac{c_1}{F_0}$. These statements all hold true for the fluxes also.

Allowing for D brane and O plane behaviour at the boundaries of the interval as in \eqref{eq:boundaries}, as well as composite objects, we find that it is possible to realise the following physical boundary behaviours
\beq\nn
\begin{array}{c|c|c|c|c}
\text{Source} & \text{Minimal tuning} & \text{M}^{1,p} & \tilde{\text{B}}^{s} & \text{Loci}  \\[2mm]
\hline
\text{Smeared D4} &c_3=0     & \text{AdS}_3\times  \text{S}^2& \text{CY}_2& \rho= \frac{\pm c_1- c_4}{c_5 \mp F_0}\\
\text{Smeared D2} & \text{Generic}~ c_i &\text{AdS}_3&\text{CY}_2\times  \text{S}^2& \rho=\frac{\pm c_1- c_4}{c_5 \mp F_0}\\
\text{D2 inside D6}&\text{Generic}~ c_i& \text{D6: }\text{AdS}_3\times \text{CY}_2& \text{D2: }\text{CY}_2& \rho= -\frac{c_2}{c_3}\\
\text{O2 inside O6}& c_1= b c_4,~ F_0= b c_5 &  \text{O6: }\text{AdS}_3\times \text{CY}_2& \text{O2: }\text{CY}_2\times  \text{S}^2& \rho= -\frac{c_4}{c_5}\\
\text{T/NATD hybrid} & \eqref{eq:TNATdualhybrid2}     & -& -& -\\
\end{array}
\eeq
where $b$ are arbitrary constants and we include the T-dual/non-Abelian T-dual hybrid, which is regular, for completeness. As before, one can also interpret the D branes smeared on all their compact co-dimensions as smeared O planes. 

As in section \eqref{eq:classICYfoliations}, we need two of these boundary behaviours to exist for the same tuning of $c_i$ to realise a compact local solution beyond the $F_0=0$ limit of \eqref{eq:hybrid2}. We find the following possibilities
\beq\label{eq:boundary behaviours4}
\begin{array}{c|c|c}
\text{Tuning} & \text{Boundary behaviours}& \text{Loci:~}\rho=\\[2mm]
\hline
\text{generic}~c_i& \text{D2}~|~ \text{D2 in D6}~|~ \text{D2} &  \frac{+ c_1- c_4}{c_5-  F_0}~|~-\frac{c_2}{c_3}~|~\frac{- c_1- c_4}{c_5+  F_0}\\
(c_4= b c_1,c_5= b F_0,c_3\neq 0) & \text{D2 in D6}~|~ \text{O2 in O6} & -\frac{c_2}{c_3}~|~-\frac{c_1}{F_0}\\
(c_2= b c_1,c_3= b F_0,c_3\neq 0) & \text{D2}~|~\text{D2} &-\frac{c_1}{F_0}~|~-\frac{c_4}{c_5}\\
c_3=0 &\text{D4}~|~\text{D4} & \frac{c_1-c_2}{c_5-F_0}~|~\frac{-c_1-c_2}{c_5+F_0}\\
\end{array}
\eeq
Together with \eqref{eq:boundary behaviours2} this gives a total of 13 distinct foliations of AdS$_3\times$S$^3\times$ CY$_2$ over intervals bounded between a rich variety of D brane and O plane behaviours. They are compact whenever CY$_2= \text{T}^4$ or K3, which really doubles the number of distinct solutions to 26.\\
~\\
As was true of section \eqref{eq:classICYfoliations}, the general solution of this section is only local. One can actually construct more general globally compact solutions by glueing these local solutions together with defect branes. In the next section we will explore this possibility. 

\section{Glueing local solutions together with defect branes}\label{sec:calibrations}
In sections \ref{eq:classICYfoliations} and \ref{eq:classIICYfoliations} we found several local compact solutions that are foliations of AdS$_3\times \text{S}^2\times \text{CY}_2$ over a finite interval bounded by various D brane an O plane behaviours. In this section we show that these compact local solutions, and more generally any local solution in these classes, may be used as the building blocks of a far larger class of globally compact solutions. This can be achieved by using  defect branes to glue the various local solutions together. This follows  the spirit of \cite{Apruzzi:2013yva}, where an infinite family of globally compact AdS$_7$ solutions in massive IIA was found, that utilised D8 brane defects to glue various non compact local solutions together (see also \cite{Macpherson:2018mif} for an AdS$_3$ example).\\
~\\
Through out most of this paper we have derived our various classes of solution under the assumption that we are in a region of the internal space away from the loci of sources. This was actually sufficient to find solutions with sources on the boundary of the internal space - as then one can explicitly see known brane/plane behaviour appearing in the physical fields.  However to realise defect branes, that lie on the interior of the internal space, we will have to explicitly solve the source corrected Bianchi identities and make sure the sources are supersymmetric.  

Various types of defect branes are possible in supergravity, with various signatures - the most simple is probably the D8. The singularity signalling a  D8 brane defect is rather mild, giving rise only to a discontinuity in the derivatives of the metric and dilaton, with the fields themselves continuous. The NS 2-form on the other hand need only be continuous up to a large gauge transformation. The remaining fluxes can be discontinuous across such a defect provided that this is induced by a shift in the D8 brane flux $F_0$ - which should naturally shift as one crosses a D8 brane stack. In what follows this will be one defect we use to perform glueings. The others are a D4 brane defect and a D6 defect that are both smeared over their compact co-dimensions. Such objects behave in a completely analogous way to the D8 defect, indeed for CY$_2$= T$^4$ they are mapped into each other via T-duality, only now it is the charge of D4s/D6s rather than $F_0$ that experiences a discontinuity as we cross the defect.\\
~\\
Having set the scene, it will now be helpful to look at the two cases individually to show that such glueing of local solutions is possible. Let  us first look at global solutions following from section \ref{eq:classICYfoliations}.

\subsection{Towards global solutions with defects from section \ref{eq:classICYfoliations}}
In this section we will study the possibility of gluing the local solutions of section \ref{eq:classICYfoliations} together with defect branes.\\
~\\
As explained in section \ref{eq:classICYfoliations}, the general local form of the NS sector and RR Page fluxes are given exactly by \eqref{eq:NSfolationI} and \eqref{eq:RRfolationI} respectively. However these expressions depend on constants $(c_i,F_0)$ that can change as we cross a defect brane, so for a global solution it is more helpful to consider the form of NS sector given in \eqref{eq:classI NS},  we remind the reader that here we fix $H_2=0$ so as to respect the symmetry of CY$_2$ and $(h_8,h_4,u)$ are all functions of $\rho$ only, the latter being linear and the behaviour of the former two determined by the Bianchi identities of the fluxes. As we shall see $(h_8,h_4)$  end up being piece-wise linear so that the ``constants''  
\beq
 F_0=h_8',~~~~G_0=h_4' 
\eeq
are not globally defined, but can change between local patches of a global solution. For the RR sector it will be most useful to know the magnetic component of the Page flux polyform
\beq
\hat f=h_8'-\frac{1}{2}(h_8-\rho h_8')\text{vol}(\text{S}^2)-\bigg(h_4'- \frac{1}{2}(h_4-\rho h_4')\text{vol}(\text{S}^2)\bigg)\wedge \text{vol}(\text{CY}_2).
\eeq
Recall the Page flux is defined in terms of the NS 2-form $B$ as $\hat{F}= e^{-B}\wedge F$, for simplicity we do not consider large gauge transformations in $B$ - however we stress that their inclusion changes nothing substantive about what follows.~\\
\\
Let us first consider a single D8 brane defect: The Bianchi identity of the entire magnetic  flux in the presence of a generic D8 brane stack takes the form
\beq\label{eq:BIforD8stack}
(d-H)f = \frac{n_8}{2\pi} \delta(\rho-\rho_0) e^{{\cal F}}\wedge d\rho 
\eeq
where $\rho_0$ is the loci of the stack, and $n_8$ its charge. As usual ${\cal F}= B+ 2\pi \tilde{ f}_2$ for $\tilde{ f}_2$ a world-volume flux that may be turned on - this should not to be confused with the RR 2-form!  As $B\sim \text{vol}(\text{S}^2)$ for the local solutions of section \ref{eq:classICYfoliations}, we anticipate that this D8 brane is actually (at least) a D8-D6 bound state - however exactly what branes are bounded together will depend on the form of $\tilde{ f}_2$ that we determine by actually solving \eqref{eq:BIforD8stack}. Following \cite{Apruzzi:2013yva}, we do this in terms of $\hat f$, for which \eqref{eq:BIforD8stack} is equivalent to
\beq\label{eq:D8pagefluxbicaseI}
d\hat f = \frac{n_8}{2\pi} \delta(\rho-\rho_0) e^{2\pi \tilde{f}_2}\wedge d\rho 
\eeq
As we move across this defect the NS sector \eqref{eq:NSfolationI}  should be continuous ($B$ can shift by a large gauge transformation, but for simplicity we shall assume it does not), while only $F_0$ should shift. Thus $h_8,h_4,h_4',h_4''$ should be continuous across the defect while $F_0=h_8'$ will be discontinuous.
As such integrating \eqref{eq:D8pagefluxbicaseI} across the D8 stack gives rise to
\beq
\Delta F_0 e^{\frac{1}{2}\rho_0\text{vol}(\text{S}^2)}= \frac{n_8}{2\pi}e^{2\pi \tilde{f}_2}
\eeq
for $\Delta F_0$ the difference between the values of $F_0$ for $\rho<\rho_0$ and  $\rho>\rho_0$. We thus see that the Bianchi identity merely fixes
\beq
\Delta F_0 = \frac{n_8}{2\pi},~~~~ \tilde{f}_2= \frac{1}{4 \pi}\rho_0\text{vol}(\text{S}^2),~~~\Rightarrow~~~ {\cal F}= \frac{ u u'}{8 h_4 h_8+ 2(u')^2}\text{vol}(\text{S}^2)
\eeq
confirming that the defect is actually a D8-D6 bound state.\\
~\\
For the D4 brane defect wrapped on AdS$_3\times$S$^2$ and smeared over CY$_2$ things are rather similar. The Bianchi identity of such a D4 brane stack takes the form
\beq\label{eq:D4pagefluxbicaseI}
(d-H)f\bigg\lvert_{\text{CY}_2} = (2\pi)^3n_4 \delta(\rho-\rho_0) e^{{\cal F}}\wedge d\rho \wedge \text{vol}(\text{CY}_2)
\eeq
where $(\rho_0,n_4)$ are the loci and  charge of the stack - the notation on the LHS means we only consider the components parallel to $\text{vol}(\text{CY}_2)$. This time it should be only $G_0=h_4'$ which is discontinuous across the defect, so integrating the Page form avatar of \eqref{eq:D4pagefluxbicaseI} gives rise to
\beq
\Delta G_0 e^{\frac{1}{2}\rho_0\text{vol}(\text{S}^2)}= -(2\pi)^3n_4e^{2\pi \tilde{f}_2},
\eeq 
where the volume of CY$_2$ has been factored out of both sides of this expression. We need then only fix
\beq
\Delta G_0 =- (2\pi)^3n_4,~~~~ \tilde{f}_2= \frac{1}{4 \pi}\rho_0\text{vol}(\text{S}^2),~~~\Rightarrow~~~ {\cal F}= \frac{ u u'}{8 h_4 h_8+ 2(u')^2}\text{vol}(\text{S}^2)
\eeq
for the Bianchi identity to be solved - which implies that like the D8, the D4 is also a bound state, this time D4-D2.\\
~\\
We have shown that both D8 and smeared D4 brane defects can be placed at arbitrary points along the interval of the AdS$_3\times$S$^2\times$CY$_2$ foliations in section \ref{eq:classICYfoliations} and still solve the source corrected Bianchi identities - provided they come as part of a bound state (D8-D6 and D4-D2). To guarantee that we actually have a solution at these loci however the branes must have a supersymmetric embedding - then supersymmetry is preserved on the defects and the remaining EOM are implied \cite{Prins:2013wza}. A major advantage of the approach we took to constructing solutions in section \ref{sec:geometricsusy} is that it allows us to determine this in the language of generalised calibrations \cite{Lust:2010by}. This is relatively simple for us because the fundamental object of this approach is the 7d bi spinors already given in \eqref{eq:specificforms1}-\eqref{eq:specificforms2}. A D brane source extended along AdS$_3$ is supersymmetric if it obeys a calibration condition - namely that DBI Lagrangian $\mathcal{L}_{DBI}= d\xi^de^{-\Phi}\sqrt{-\det(g+ \cal{F})}$ is equal to a calibration form.  In IIA this calibration form is given by the pull back of  $e^{3A-\Phi}\text{vol}(\text{AdS}_3)\wedge \Psi_+\wedge e^{{\cal F}}$ onto the relevant D brane world-volume. It is  not hard to show that both our D8 and D4 brane defects obey this condition precisely when  ${\cal F}$ is tuned as the Bianchi identity of each defect requires.\\
~\\
Thus we have established that one can place defects at arbitrary points along the interval of the  AdS$_3\times$S$^2\times$CY$_2$ foliation and still have a supersymmetric solution - we need only impose that $(h_4,h_8)$ are continuous. This fact can be used to glue two local solutions of section \ref{eq:classICYfoliations} together provided they share a common tuning for $u$ ($u''=0$ by supersymmetry, so $u$ is globally linear). There is no limit to the number of defects one can place in a global solution, indeed in general $(h_4,h_8)$ need only be piece-wise linear with a change in slope of the former (latter) indicating the presence of a D4 (D8) brane at that loci. One can therefore construct infinite classes of global solutions for each tuning of $u$ in section \ref{eq:classICYfoliations}. We delay a detailed exploration of these possibilities and their interpretation in terms of the AdS$_3/$CFT$_2$ correspondence  until
\cite{LMNR3,LMNR2}\\
~\\
In the next section, we explore the possibility of constructing global solutions with defects from the solutions in section \ref{eq:classIICYfoliations}

\subsection{Towards global solutions with defects from section \ref{eq:classIICYfoliations}}
In this section we will show that it is possible to glue the local solutions of section  \ref{eq:classIICYfoliations} together with defect branes. As most of the details of this procedure are covered in the previous section, we encourage the reader to go over that first, as here we shall be brief.\\
~~\\
As before the local solutions of section \ref{eq:classIICYfoliations} depend on constants $(c_i,F_0)$.  However, as these constants can shift between local patches in a global solution, it is more helpful to consider the NS sector in the form  of \eqref{eq:classIINS}, with $h=1$ and $\hat{\text{M}}_4=\text{CY}_2$ - recall $(u,v,w)$ are functions of $\rho$ only and that $u$ is such that globally $u''=0$ due to supersymmetry.  Conversely $v,w$ are only linear functions away from localised sources - globally they need only be piecewise linear provided that the resulting $\delta$-functions appearing in their second derivatives  gives rise to a source corrected Bianchi identity, and this source is calibrated. The magnetic Page flux polyform associated to these solutions is
\begin{align}
\hat f&= v'-\frac{1}{2}(v-\rho v')\text{vol}(\text{S}^2)- \big(w'-\frac{1}{2}(w-\rho w'\big)\text{vol}(\text{S}^2)\big)\wedge  \hat J\nn\\[2mm]
&+ \big(v'-\frac{1}{2}(v-\rho v')\text{vol}(\text{S}^2)\big)\wedge \text{vol}(\text{CY}_2)\label{eq:hatfcalssII}.
\end{align}
Given the form of this expression it might be tempting to interpret a shift in $v'$  as coincident D8-D6 and D4-D2 bound states - however such configurations fail to obey the calibration condition discussed in the previous section at generic points along the interval - so cannot be used to glue solutions together without breaking supersymmetry\footnote{They do obey the calibration condition when $w=v$, where the metric blows up, so they still cannot be interpreted as defect branes, like in the previous section.}.  Shifts in $w'$ on the other hand are different and give rise to something new. To interpret it consider the following: If we take a D8-D6 brane defect wrapping CY$_2=$T$^4$, we can express $\hat J= dx_1\wedge dx_2+ dx_3\wedge dx_4$ with $x_i$ the directions on T$^4$ which are all isometries. If one T-dualises such an object on both $(x_1,x_2)$ it would generate the part of $- \big(w'-\frac{1}{2}(w-\rho w'\big)\text{vol}(\text{S}^2)\big)\wedge  \hat J$ with legs in $(x_1,x_2)$ - this is a D6-D4 brane wrapping AdS$_3\times$S$^2$ and $(x_3,x_4)$ which is smeared on  $(x_1,x_2)$. If instead one T-dualised the D8-D6 bound state on $(x_3,x_4)$, the part of the previous expression with legs in $(x_1,x_2)$ would be generated, which should be interpreted as a D6-D4 wrapping $(x_1,x_2)$ and smeared on $(x_3,x_4)$. To generate the entire  $w$
dependent term in \eqref{eq:hatfcalssII} then, one should have both of these smeared D6-D4s simultaneously. Generalising to generic CY$_2$, a shift in $w'$ gives rise to a D6-D4 bound state that wraps a curve in CY$_2$ and is smeared over its co-cycle and another D6-D4 that wraps and is  smeared over the opposite cycles. The Bianchi identities of each bound state are essentially the same as the D4-D2 of the previous section, only this time pulled back onto the relevant curve rather than the entire of CY$_2$ - they are solved as before with world volume gauge field $4\pi \tilde{f}_2=\rho_0\text{vol}(S^2)$ and D6 brane charge  proportional to $\Delta w'$ across the defect. Finally, it is not hard to establish that each of  the D6-D4 bounds states are indeed calibrated at generic points in the space.\\
~\\
We have now established that D6-D4 defect branes can be placed at generic points along the interval of the  AdS$_3\times$S$^2\times$CY$_2$ foliation of section \ref{eq:classIICYfoliations}. It would be interesting to explore what global solutions may be constructed by glueing the local solutions already found together with these defects. We leave that for future work.\\
~\\
In the next and final section we summarise this work and discuss some future directions.
\section{Summary and future directions}\label{summery}
In this paper we have found two classes of warped AdS$_3\times  \text{S}^2\times \text{M}_5$ solutions in massive IIA that preserve small $\mathcal{N}=(4,0)$ supersymmetry in terms of an SU(2) structure on  $\text{M}_5$. These classes are exhaustive for solutions of this type when one assumes that the associated spinors on $ \text{S}^2\times \text{M}_5$ have equal norm, a requirement for non vanishing Romans mass. For class I  $\text{M}_5$ decomposes as $\text{CY}_2\times \mathbb{R}$ and we are able to give explicit local expressions for the metric and fluxes up to simple Laplace like PDEs. This class contains a generalisation of the flat space system of D4s inside the world volume of D8s contained in \cite{Youm:1999zs}, with flat space  replaced by AdS$_3\times  \text{S}^2\times \text{CY}_2\times \mathbb{R}$. For class II we find $\text{M}_5=\text{M}_4\times \mathbb{R}$ where $\text{M}_4$ is a class of warped Kahler manifolds with metrics that depend on the interval.\\
~\\
Performing T duality on the IIA classes, we find new classes of solutions in IIB that, modulo SL$(2,\mathbb{R})$ transformations, exhaust  $\mathcal{N}=(4,0)$ solutions of the type AdS$_3\times \text{S}^3\times \text{M}_4$, with $\text{M}_4$ an SU(2) structure manifold. The first is a generalisation of the near horizon limit of D1-D5 branes \cite{Maldacena:1997re}, where the $\text{S}^3$ becomes fibered over CY$_2$ and D5 branes are backreacted on top of this. It is possible to turn off the fibre and then realise the resulting system as a near horizon limit with a modification of the D1-D5 intersection. The second class of IIB solutions is a generalisation of D3 branes wrapped on a curve inside the base of an elliptically fibered CY$_3$ \cite{Couzens:2017way}. The generalisation depends on the same necessary geometric conditions as  \cite{Couzens:2017way}, but has an additional parameter turned on which is related to the charge of 5-branes, absent in the original construction, which tunes the 3-forms to zero.\\
~\\
In sections \ref{eq:classICYfoliations} and \ref{eq:classIICYfoliations}, we have found several new local solutions in massive IIA that are foliations of AdS$_3\times  \text{S}^2\times \text{CY}_2$ over an interval,  bounded between a variety of D brane and O plane behaviours. Then in section \ref{sec:calibrations} we show how these may be used as the building blocks of infinite families of global solutions. These utilise defect branes to glue the various local solutions together in the vein of \cite{Apruzzi:2013yva}. We will explore some possible global solutions containing defect branes and their holographic interpretation in \cite{LMNR2,LMNR4}. \\
~\\
An interesting open problem that our classification of (0,4) supersymmetric solutions leaves is the identification of their 2d dual CFTs. On the other hand, as stressed in the introduction, there are large classes of 2d (0,4) linear quivers, such as the ones constructed in \cite{Haghighat:2013gba,Haghighat:2013tka,Gadde:2015tra,Hanany:2018hlz}, which lack a holographic description. In \cite{LMNR2} we will partially fill this gap, and provide the explicit connection between AdS$_3\times  \text{S}^2$ solutions in class I  with compact CY$_2$ and 4d (0,4) quivers.\\
~\\ 
Another interesting avenue to explore as a consequence of this work is the connection between our solutions and the AdS$_7$ solutions to massive IIA constructed in \cite{Apruzzi:2013yva}, in particular whether a generalisation to AdS$_3$ solutions exists of the flows constructed in \cite{Apruzzi:2015wna,Apruzzi:2015zna,Rota:2015aoa}. We will report progress in this direction in \cite{LMNR4,LMNR3}.

~\\


\subsection*{Acknowledgements}
We would like to thank  Christopher Couzens, Giuseppe Dibitetto, Nicolo Petri, Daniel Prins,  Alessandro Tomasiello and Stefan Vandoren for fruitful discussions. YL and AR are partially supported by the Spanish government grant PGC2018-096894-B-100 and by the Principado de Asturias through the grant FC-GRUPIN- IDI/2018/000174.   NTM is funded by the Italian Ministry of Education, Universities and Research under the Prin project ``Non Perturbative Aspects of Gauge Theories and Strings'' (2015MP2CX4) and INFN. C.N. is Wolfson Research Merit Fellow of the Royal Society. AR is supported by CONACyT- Mexico scholarship. We would like to acknowledge the Mainz Institute for Theoretical Physics (MITP) of the DFG Cluster of Excellence PRISMA$^{+}$ (Project ID 39083149) for its hospitality and partial support during the development of this work. YL and AR would also like to thank the Theory Unit at CERN for its hospitality and partial support during the completion of this work.

\appendix

\section{Spinors and bi spinors on $ \text{S}^2$ and M$_5$}\label{sec:app}
In this appendix we provide details of the spinors and bi spinors on $ \text{S}^2$ to supplement section \ref{sec:geometricsusy}. Specifically when deriving the 7 dimensional bi spinors \eqref{eq:bispinordef} from \eqref{eq:neq1spinor}, it is useful to know the 2 and 5 dimensional bi spinors on $ \text{S}^2$ and $\text{M}_5$, which \eqref{eq:bispinordef} will decompose in terms of. In fact given our decomposition of the gamma matrices \eqref{eq:7dgammas}, a bi spinor constructed out of tensor products of spinors in 2  and 5 dimensions ($\xi^i$ and $\eta^i$ respectively) necessarily decomposes as
\beq
\big[\xi^1\otimes \eta^1\big]\otimes \big[\xi^2\otimes \eta^{2}\big]^\dag = (\eta^1\otimes \eta^{2\dag})_+\wedge (\xi^1\otimes\xi^{2\dag}) +(\eta^1\otimes \eta^{2\dag})_-\wedge (\sigma_3\xi^1\otimes\xi^{2\dag})
\eeq
where $\pm$ denotes the even/odd degree components of a form only, which can be repeatedly used when computing \eqref{eq:bispinordef} - and proves that it is built from bi spinors in 2 and 5 dimensions.\\
~\\
In the next section we present details of spinors and bi spinors on unit norm $ \text{S}^2$.  

\subsection{Spinors and bi spinors on $ \text{S}^2$}
There are two types of Killing spinor on unit radius $ \text{S}^2$, $\xi_{\pm}$, that are solutions to the Killing spinor equations
\beq
\nabla_a \xi_{\pm}= \pm \frac{i}{2}\sigma_{a}\xi_{\pm},~~~a=1,2,
\eeq
where $a$ is a flat index and we take the first 2 Pauli matrices as two dimensional flat space gamma matrices. Unlike the $\xi_{\pm}$ equivalents on $\text{S}^3$, these are not really independent and in fact one can take $\xi_{-}=\sigma_3\xi_+$ without loss of generality. We identify $\xi_+=\xi$ in the main text, and one has in general that both $\xi$ and $\sigma_3\xi$ transform in the same fashion under the SU(2) global symmetry on $ \text{S}^2$.
The bi spinors that follow  from $\xi$, under the assumption they have unit norm, are \cite{Macpherson:2016xwk}, 
\begin{align}\label{eq:S2bispinors}
\xi\otimes\xi^{\dag}&=\frac{1}{2}(1+ k_3-i y_3 \text{vol}( \text{S}^2)),\quad \xi\otimes\xi^{c\dag}=-\frac{1}{2}(k_1+ i k_2- i (y_1+ i y_2)\text{vol}( \text{S}^2)),\nn\\[2mm]
\sigma_3\xi\otimes\xi^{\dag}&=\frac{1}{2}(y_3+ i dy_3- i \text{vol}( \text{S}^2)),\quad \sigma_3\xi\otimes\xi^{c\dag}=-\frac{1}{2}( y_1+ i y_2-i d(y_1+ i y_2),
\end{align}
where $y_i$ are coordinates embedding a unit radius $ \text{S}^2$ into $\mathbb{R}^3$ and $K_i$ are one forms dual to the Killing vectors of SU$(2)$, which may be parameterised as
\beq
K_i= \epsilon_{ijk}y_j dy_k.
\eeq
Note that \eqref{eq:S2bispinors} are spanned entirely by the $(y_i,dy_i,K_i, y_i \text{vol}( \text{S}^2))$ which transform as SU(2) triplets, and $\text{vol}( \text{S}^2)$, that is an SU(2) singlet. These form a closed set under the action of $d$ and wedge product, namely
\beq\label{eq:SU2forms}
dy_i\wedge \text{vol}( \text{S}^2)=K_i\wedge \text{vol}( \text{S}^2)=0,~~~ dK_i=2y_i\text{vol}( \text{S}^2)
\eeq 
as well as the more obvious relations. We use this fact to reduce the 7d conditions that follow from inserting \eqref{eq:specificforms1}-\eqref{eq:specificforms2} into \eqref{eq:susycond7d1}-\eqref{eq:susycond7d3}  to a set of 5d conditions no longer involving $ \text{S}^2$, \eqref{eq:gensusyCOND1}-\eqref{eq:paring}.\\
~\\
In the next section we give details on the bi spinors in 5d. 

\subsection{Spinors and bi spinors on $M^5$}\label{sec:5dbispinors}
In \eqref{eq:indedidual5dsinors} we decompose the independent 5d spinors appearing in \eqref{eq:neq1spinor} in terms of a single unit norm spinor in 5d, $\eta$.
The bi-linears that follow from $\eta$ are given in \cite{Apruzzi:2015zna}, and read: 
\begin{align}
\eta\otimes \eta^{\dag}&= \frac{1}{4}(1+V)\wedge e^{-i j_3},\quad \eta\otimes \eta^{c\dag}= \frac{1}{4}(1+V)\wedge \Omega,\nn\\[2mm]
\Omega&= w\wedge u,~~~~ j_3=\frac{i}{2}(w\wedge \overline{w}+u\wedge \overline{u}),
\end{align}
where 
\beq
v,~w_1=\text{Re}w,~w_2=\text{Im}w~u_1=\text{Re}u,~u_2=\text{Im}u
\eeq
defines a vielbein in five dimensions. It then follows that if one decomposes
\beq
\Omega= j_1+ i j_2,
\eeq
we have 
\beq
j_a\wedge j_b= \frac{1}{2}\delta_{ab}\text{vol}(\text{M}_4)
\eeq
where $V\wedge \text{vol}(\text{M}_4)$ is the volume form in 5d.

\end{document}